\documentclass[
 amsmath,amssymb,
 aps,
 reprint,
pra
]{revtex4-2}

\usepackage{graphicx}
\usepackage{dcolumn}
\usepackage{bm}
\usepackage{physics}
\usepackage{amsfonts, amsmath}
\usepackage{dsfont}
\usepackage{amsthm}
\usepackage{bm}
\usepackage{hyperref}
\usepackage{bbold}
\usepackage{color}
\usepackage{lipsum,babel}
\usepackage[normalem]{ulem}
\usepackage{tabularx}
\usepackage{comment}
\usepackage{float}
\makeatletter
\let\newfloat\newfloat@ltx
\makeatother
\usepackage{algorithm}
\usepackage{algpseudocode}

\newcommand{\cor}[1]{{\color{red}{#1}}}

\newtheorem{theorem}{Theorem}

\newtheorem*{theorem-restate}{Theorem \ref{theorem:bounding ere}}

\begin{document}
\definecolor{navy}{RGB}{46,72,102}
\definecolor{pink}{RGB}{219,48,122}
\definecolor{grey}{RGB}{184,184,184}
\definecolor{yellow}{RGB}{255,192,0}
\definecolor{grey1}{RGB}{217,217,217}
\definecolor{grey2}{RGB}{166,166,166}
\definecolor{grey3}{RGB}{89,89,89}
\definecolor{red}{RGB}{255,0,0}

\preprint{APS/123-QED}

\title{Matrix product state approach to lossy boson sampling and noisy IQP sampling}

\author{Sojeong Park}
\affiliation{Department of Physics, Korea Advanced Institute of Science and Technology, Daejeon 34141, Korea}

\author{Changhun Oh}
\email{changhun0218@kaist.ac.kr}
\affiliation{Department of Physics, Korea Advanced Institute of Science and Technology, Daejeon 34141, Korea}

\begin{abstract}
Sampling problems have emerged as a central avenue for demonstrating quantum advantage on noisy intermediate-scale quantum devices. However, physical noise can fundamentally alter their computational complexity, often making them classically tractable. Motivated by the recent success of matrix product state~(MPS)-based classical simulation of Gaussian boson sampling~(Oh et al., 2024), we extend this framework to investigate the classical simulability of other noisy quantum sampling models. We develop MPS-based classical algorithms for lossy boson sampling and noisy instantaneous quantum polynomial-time~(IQP) sampling, both of which retain the tunable accuracy characteristic of the MPS approach through the bond dimension. Our approach constructs pure-state decompositions of noisy or lossy input states whose components remain weakly entangled after circuit evolution, thereby providing a means to systematically explore the boundary between quantum-hard and classically-simulable regimes. For boson sampling, we analyze single-photon, Fock, and cat-state inputs, showing that classical simulability emerges at transmission rates scaling as 
$O(1/\sqrt{N})$, reaching the known boundary of quantum advantage with a tunable and scalable method. Beyond reproducing previous thresholds, our algorithm offers significantly improved control over the accuracy–efficiency trade-off. It further extends the applicability of MPS-based simulation to broader classes of noisy quantum sampling models, including IQP circuits.
\end{abstract}

\maketitle{}

\section{Introduction}

It is widely believed that a quantum computer can efficiently solve certain problems that are hard for classical computers, such as integer factoring and Hamiltonian simulation~\cite{shor1994algorithms,lloyd1996universal}.
Since one of the most critical obstacles to realizing a quantum computer is noise, quantum error correction~(QEC) has been developed to actively detect and correct errors, enabling fault-tolerant quantum computation~\cite{shor1995scheme,steane1996error}.
Despite substantial advances in both quantum hardware and the theory and implementation of QEC~\cite{google2025quantum,terhal2015quantum}, currently available quantum devices cannot yet implement QEC in a fully scalable manner.
Consequently, extensive efforts have focused on achieving quantum advantage using currently available quantum devices, often referred to as noisy intermediate-scale quantum~(NISQ) devices~\cite{aaronson2011computational, bouland2019complexity, arute2019quantum, wu2021strong, morvan2023phase, zhong2020quantum, zhong2021phase, madsen2022quantum, deng2023gaussian, young2024atomic, liu2025robust}.

To pursue quantum advantage with NISQ devices, quantum sampling problems have emerged as promising candidates due to their complexity-theoretic hardness results in noiseless cases and their relatively feasible experimental requirements, compared with more sophisticated quantum algorithms~\cite{aaronson2011computational, bouland2019complexity, arute2019quantum}.
The most representative examples of sampling problems include boson sampling, random circuit sampling, and instantaneous quantum polynomial-time~(IQP) sampling~\cite{aaronson2011computational,boixo2018characterizing,bremner2011classical}.
Indeed, several experiments based on these sampling problems have reported evidence for quantum advantage~\cite{wang2019boson,zhong2020quantum,madsen2022quantum,arute2019quantum,bluvstein2024logical,liu2025robust}.


However, such claims have been continuously challenged by the presence of noise in realistic experiments (e.g., Refs.~\cite{neville2017classical,clifford2018classical,oszmaniec2018classical,garcia2019simulating,qi2020regimes,oh2021classical,quesada2022quadratic,bulmer2022boundary,oh2022classical,oh2023classical,liu2023simulating,oh2024classical, oh2025recent, oh2025classical}).
The underlying reason is that physical noise can degrade, and in some regimes destroy, the computational advantage.
For example, in the presence of depolarizing noise in quantum circuits and at large circuit depth, the system accumulates entropy, causing the output state to converge to the maximally mixed state, which is easily simulated classically~\cite{aharonov1996limitations}.
Similarly, in photonic systems, it has been shown that photon loss makes a quantum state converge to a thermal state, which is easy to classically simulate~\cite{garcia2019simulating,qi2020regimes}.
As such, physical noise can move sampling problems from a regime believed to be classically intractable to one that admits efficient classical simulation.
Hence, to rigorously assess the potential advantage of quantum devices, it is essential to understand whether and when sampling problems become easy to classically simulate under realistic noise conditions.

Boson sampling provides a particularly well-studied example of this phenomenon: when photon loss is sufficiently strong, the problem becomes easier to simulate classically. 
Motivated by this observation, a variety of classical simulation algorithms have been developed~\cite{rahimi2016sufficient, garcia2019simulating, huang2019simulating, clifford2020faster, oh2021classical,liu2023simulating}. 
One line of work~\cite{garcia2019simulating} used a strategy to approximate a lossy quantum state by a quantum state that is classically easy to simulate.
For example, Ref.~\cite{oszmaniec2018classical} and Ref.~\cite{garcia2019simulating} showed that a lossy single photon state can be approximated by a separable state or a thermal state, respectively, when the loss rate is high.
Furthermore, this idea has been extended to lossy Gaussian boson sampling~(GBS)~\cite{qi2020regimes}.
Although these approximations provide important benchmarks, they were not sufficient to simulate recent large-scale GBS experiments~\cite{madsen2022quantum,zhong2020quantum}.
A key limitation of these approaches is that, since they rely on approximating a given state by the nearest classically simulable state, their performance cannot be systematically improved by allocating additional computational resources.
Thus, outside the high-loss regime, these fixed approximations do not provide a systematic route to improve the simulation accuracy.

A more recent MPS-based algorithm overcame this limitation for lossy GBS and enabled the simulation of the largest GBS experiments available at the time.
Its crucial feature is tunability: increasing the MPS bond dimension systematically improves the simulation accuracy at the cost of additional computation.
However, the construction in Ref.~\cite{oh2024classical} relies strongly on Gaussian structure, and its extension to non-Gaussian sampling models has remained unclear.


In this work, we extend the decomposition-based MPS strategy beyond GBS. The key idea is to express a noisy input state as a probabilistic mixture of pure product states whose entanglement remains controlled under the relevant circuit family. 
For lossy boson sampling, we construct such decompositions for single-photon inputs, multiphoton Fock inputs, and cat-state inputs.
For the single-photon case, we prove a R\'enyi-entropy bound that is uniform over arbitrary passive interferometers and arbitrary output-mode bipartitions. 
This gives an efficiently MPS-approximable regime approaching the familiar $O(1/\sqrt{N})$ transmission scaling found in previous works on lossy boson sampling and GBS~\cite{garcia2019simulating, oszmaniec2018classical,qi2020regimes,oh2024classical}, while retaining a tunable accuracy--cost tradeoff through the MPS bond dimension. We further apply the same principle to noisy IQP circuits with dephasing and depolarizing noise and obtain a noise-depth threshold comparable to recent results derived by different methods~\cite{nelson2024polynomial}.

The remainder of this paper is organized as follows. In Sec.~\ref{sec:probsetup}, we introduce the problem setup by describing boson sampling and IQP sampling circuits and provide the relation between MPS and efficient classical simulation in Sec.~\ref{sec:MPS}. In Sec.~\ref{sec:BS}, we present an MPS-based classical simulation algorithm for boson sampling with various input states and analyze the asymptotically simulable range and numerically estimate computational resources. In Sec.~\ref{sec:IQP}, we provide a classical simulation algorithm for IQP sampling together with its numerical analysis. Finally, Sec.~\ref{sec:discussion} discusses the implications of our results and concludes the paper.

\section{Problem setup}
\label{sec:probsetup}
\subsection{Boson sampling}
\label{sec:BSsetup}
Boson sampling consists of an $M$-mode passive linear-optical interferometer with $N$ single photons injected into the first $N$ modes and vacuum in the remaining modes~\cite{aaronson2011computational}. Let $\hat a_i$ and $\hat a_i^\dagger$ denote the annihilation and creation operators of the $i$th mode, satisfying $[\hat a_i,\hat a_j^\dagger]=\delta_{ij}$ and $[\hat a_i,\hat a_j]=0$. The input state is
\begin{align}
|1\rangle^{\otimes N}|0\rangle^{\otimes(M-N)}
=
\left(\prod_{i=1}^N \hat a_i^\dagger\right)|0\rangle^{\otimes M}.
\end{align}

The interferometer is specified by an $M\times M$ unitary matrix $U$. In the Heisenberg picture, the creation operator of the $j$th input mode transforms as
\begin{align}
\hat a_j^\dagger
\mapsto
\sum_{k=1}^M U_{jk}^* \hat a_k^\dagger ,
\end{align}
where the operators on the right-hand side are the physical output-mode creation operators. After the interferometer, photon-number-resolving measurement in these output modes produces an occupation pattern $\vec t=(t_1,\ldots,t_M)\in\mathbb{Z}_{\ge 0}^M$ satisfying $\sum_{i=1}^M t_i=N$.
For an ideal interferometer, the probability of observing an output pattern $\vec t$ is
\begin{align}
\Pr(\vec t)
=
\frac{|\operatorname{Per} U_T|^2}{t_1!\cdots t_M!},
\end{align}
where $U_T$ is obtained from the first $N$ columns of $U^\dagger$ by repeating the $j$th row $t_j$ times. This permanent structure underlies the standard complexity-theoretic evidence that ideal boson sampling is hard to simulate classically.

We model photon loss by coupling each optical mode to an environmental vacuum mode through a beam splitter of transmission rate $\eta$: $\hat{a}_i^\dagger \to \sqrt{\eta}\hat{a}_i^\dagger +\sqrt{1-\eta}\hat{e}_i^\dagger,$ where $\hat{e}_i^\dagger$ is the creation operator of the corresponding environmental mode, which is traced out.
Under this channel, a single-photon state transforms as
\begin{align}
|1\rangle\langle 1|\rightarrow (1-\eta)|0\rangle\langle 0|+\eta|1\rangle\langle 1|.
\end{align}

For most of the analysis, we assume uniform loss, so that all modes have the same total transmission rate $\eta$. Under this assumption, loss channels commute with passive linear-optical elements and can be moved to the input of the ideal interferometer, as shown in Fig.~\ref{fig1:setup_figure}(a)--(b). Nonuniform loss can also be handled by progressively commuting the loss channels to the input, with only polynomial overhead in the number of modes and optical elements~\cite{brod2020classical}.


\subsection{IQP sampling}
\label{sec:IQPsetup}

An $n$-qubit IQP circuit prepares $|+\rangle^{\otimes n}$, applies a depth-$d$ circuit composed of gates diagonal in the computational basis, and measures all qubits in the $X$ basis. Equivalently, after the diagonal circuit, the state has the form
\begin{align}
\frac{1}{\sqrt{2^n}}
\sum_{\vec{x}\in\{0,1\}^n}
e^{if(\vec{x})}|\vec{x}\rangle,
\end{align}
where $f$ is a real-valued function determined by the diagonal gates. The final $X$-basis measurement can be implemented by applying Hadamard gates followed by computational-basis readout.



As in boson sampling, noise can make IQP sampling easier to simulate classically~\cite{bremner2017achieving,nelson2024polynomial,rajakumar2025polynomial}. 
We study this effect for Pauli noise channels,
\begin{align}
&\mathcal{N}_{p_X,p_Y,p_Z}(\hat{\rho}) \nonumber \\
&= 
(1-p_X-p_Y-p_Z)\hat{\rho}
+p_X \hat{X}\hat{\rho}\hat{X}
+p_Y \hat{Y}\hat{\rho}\hat{Y}
+p_Z \hat{Z}\hat{\rho}\hat{Z},
\end{align}
where $p_X$, $p_Y$, and $p_Z$ are the corresponding Pauli error probabilities. 
We focus on two representative cases, dephasing noise and depolarizing noise.

\begin{figure}[t]
    \centering
    \includegraphics[width=\columnwidth]{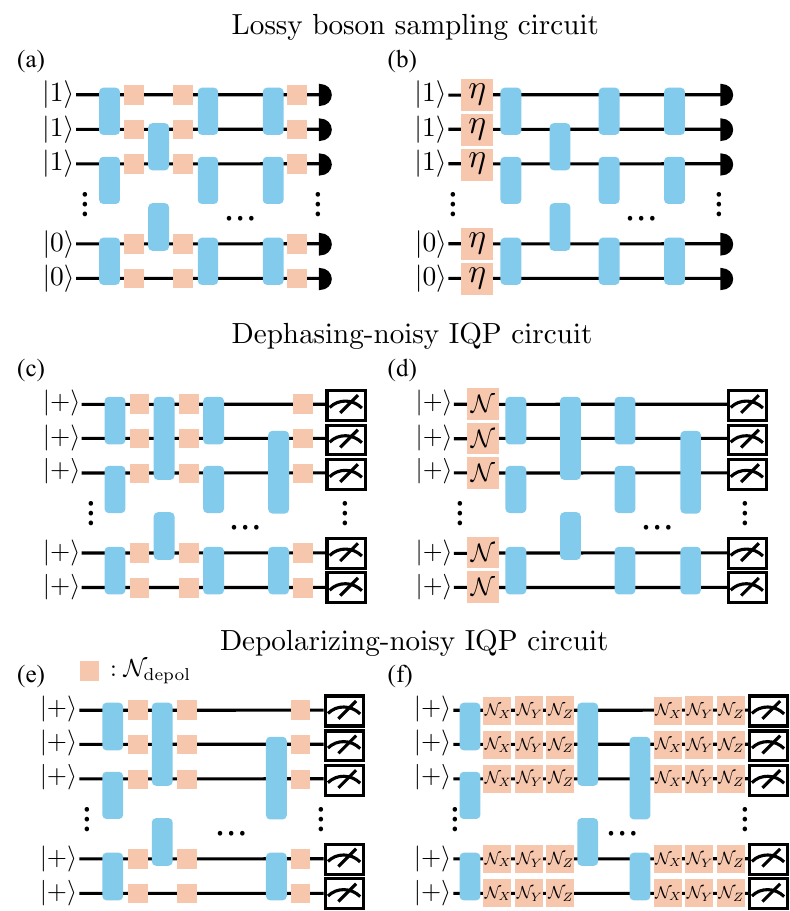}
    \caption{Mathematically equivalent noisy circuits used in this work. 
    (a)--(b) Setup of a lossy boson sampling circuit. Uniform photon loss in boson sampling is commuted to the input of the passive interferometer. Blue and orange blocks denote ideal linear optics and photon loss channels, respectively.
    (c)--(d) Setup of a dephasing-noisy IQP circuit. Dephasing noise~(orange blocks) in an IQP circuit is commuted to the input and combined into $\mathcal{N}_{0,0,p_d}$. 
    (e)--(f) Setup of a depolarizing-noisy IQP circuit. Each depolarizing channel is decomposed into $X$-, $Y$-, and $Z$-error channels, each with rate $q$. Blue and orange blocks denote ideal gates and noise channels, respectively.}
    \label{fig1:setup_figure}
\end{figure}

For dephasing noise, $p_X=p_Y=0$, and
\begin{align}
\mathcal{N}_{0,0,p}(\hat{\rho})
=
(1-p)\hat{\rho}
+
p\hat{Z}\hat{\rho}\hat{Z},
\end{align}
with $0\leq p\leq 1/2$. 
Because both this noise channel and the IQP gates are diagonal in the computational basis, the dephasing channels commute with the circuit gates and can be moved to the input, as illustrated in Fig.~\ref{fig1:setup_figure}(c)--(d). 
If the same dephasing rate $p$ is applied at each of the $d$ layers, the accumulated input noise is $\mathcal{N}_{0,0,p_d}$ with $p_d=[1-(1-2p)^d]/2.$

We also consider depolarizing noise, defined in our parametrization as
\begin{align}
\mathcal{N}_{\rm depol}(\hat{\rho})
=
\left(1-\frac{3p}{2}\right)\hat{\rho}
+
\frac{p}{2}
\left(
\hat{X}\hat{\rho}\hat{X}
+
\hat{Y}\hat{\rho}\hat{Y}
+
\hat{Z}\hat{\rho}\hat{Z}
\right),
\end{align}
where $0\leq p\leq 1/2$. 
It can be decomposed into single-Pauli noise channels as
\begin{align}
    \mathcal{N}_{0,0,q}
    \circ
    \mathcal{N}_{0,q,0}
    \circ
    \mathcal{N}_{q,0,0}
    =
    \mathcal{N}_{\rm depol},
\end{align}
where $q=[1-\sqrt{1-2p}]/2.$
This representation allows us to treat depolarizing noise by combining the input-noise decomposition used for dephasing with stochastic Pauli-error sampling, as illustrated in Fig.~\ref{fig1:setup_figure}(e)--(f).

\section{MPS and classical simulability}
\label{sec:MPS}
This work uses MPS to simulate pure-state branches obtained from noisy quantum circuits. The motivation is that physical noise often suppresses entanglement, while MPS methods are efficient precisely when the relevant states have small entanglement. 
More specifically, for a pure state on $L$ sites, consider the Schmidt decomposition across the cut $[1:l]:[l+1:L]$,
\begin{align}
|\psi\rangle
=
\sum_{\nu\geq 1} s_\nu^{(l)}
|L_\nu^{(l)}\rangle |R_\nu^{(l)}\rangle ,
\end{align}
where the Schmidt coefficients are ordered in nonincreasing order. A bond-$\chi$ MPS keeps at most the largest $\chi$ Schmidt coefficients across each cut. Increasing $\chi$ improves the approximation accuracy but also increases the computational cost. Thus, the efficiency of an MPS simulation depends on how rapidly the Schmidt coefficients decay across the relevant bipartitions.

The decay of the Schmidt coefficients can be quantified by the R\'enyi entanglement entropy. For a density operator $\hat{\rho}$ and $\alpha\geq 0$, $\alpha\neq 1$, the R\'enyi entropy is defined as
\begin{align}
S_\alpha(\hat{\rho})
\equiv
\frac{1}{1-\alpha}
\log \operatorname{Tr}(\hat{\rho}^{\alpha}),
\end{align}
with the von Neumann entropy recovered in the limit $\alpha\to 1$. For a pure state $|\psi\rangle$ on $L$ sites, its R\'enyi entanglement entropy across the cut $[1:l]:[l+1:L]$ is
\begin{align}
S_\alpha^l(|\psi\rangle)
\equiv
S_\alpha\left(
\operatorname{Tr}_{[l+1:L]}
|\psi\rangle\langle\psi|
\right).
\end{align}
For fixed $0<\alpha<1$, a standard sufficient condition for efficient MPS approximability is that $S_\alpha^l(|\psi\rangle)=O(\log L)$ for every cut $l$~\cite{schuch2008entropy}. Under this condition, $|\psi\rangle$ admits an MPS approximation with bond dimension polynomial in $L$ and in the inverse target accuracy. We use this criterion to identify loss or noise regimes in which the sampled pure states can be efficiently simulated by MPS.

We also use the complementary direction of this criterion. If, for some $\alpha>1$ and some cut $l$, the R\'enyi entanglement entropy grows algebraically, 
$S_\alpha^l(|\psi\rangle)
=
\Omega(L^\kappa)$
with a constant $\kappa>0$, then a polynomial bond dimension cannot approximate the state across that cut to fixed accuracy~\cite{schuch2008entropy}. This obstruction is specific to the MPS approximability criterion used here. It should not be interpreted as a hardness statement against all possible classical simulation methods.

\section{Classical simulation of lossy boson sampling}
\label{sec:BS}
We now apply the MPS framework to lossy boson sampling. The central task is to choose a pure-state ensemble decomposition of the lossy input such that typical sampled branches remain weakly entangled after the passive interferometer. We first present the decomposition for lossy single-photon inputs and prove an entanglement bound for the resulting sampled branches. We then estimate the required bond dimension numerically and discuss extensions to multiphoton Fock and cat-state inputs.

\subsection{Decomposition of lossy input state}

Consider the lossy boson-sampling setup of Sec.~\ref{sec:BSsetup}, with $N$ single-photon inputs and $M-N$ vacuum inputs. 
After commuting the uniform loss channels to the input, as in Fig.~\ref{fig1:setup_figure}(b), the input state becomes
\begin{align}
\hat{\rho}_{\rm in}
=
\hat{\sigma}^{\otimes N}
\otimes
|0\rangle\langle 0|^{\otimes(M-N)},
\end{align}
where
\begin{align}
\hat{\sigma}
=
(1-\eta)|0\rangle\langle 0|
+
\eta |1\rangle\langle 1|
\end{align}
is the lossy single-photon state. 
Because $\hat{\rho}_{\rm in}$ is mixed, it cannot be directly represented as a pure-state MPS. 
We therefore use an ensemble-sampling representation of the mixed input. 
For any decomposition
$\hat{\rho}_{\rm in}
=
\sum_i p_i |\psi_i\rangle\langle\psi_i|,$
weak sampling from the output distribution can be performed by first drawing $i$ with probability $p_i$ and then simulating the pure state $\hat U|\psi_i\rangle$ by MPS. 
Before MPS truncation, this procedure exactly reproduces the mixed-state output distribution:
\begin{align}
    \Pr(x)
    =
    \Tr\!\left[
    E_x \hat U\hat{\rho}_{\rm in}\hat U^\dagger
    \right] 
    =
    \sum_i p_i
    \Tr\!\left[
    E_x\hat U|\psi_i\rangle\langle\psi_i|\hat U^\dagger
    \right],
\end{align}
where $\{E_x\}$ denotes the final measurement. 
We emphasize that this is a generative weak-simulation procedure: it produces samples from the mixed-state output distribution, rather than estimating arbitrary individual output probabilities. 
If the same ensemble were used as a trajectory estimator for a fixed probability or a rare event, its variance could be large.

The key question is, therefore, which ensemble decomposition to use. 
Different decompositions of the same mixed state can lead to very different MPS costs. 
For the lossy single-photon state, the most direct choice is the Fock-basis mixture
\begin{align}
    \hat{\sigma}
    =
    (1-\eta)|0\rangle\langle 0|
    +
    \eta |1\rangle\langle 1|.
\end{align}
This decomposition has a simple operational meaning: it samples whether each photon is lost or survives. 
However, it is not useful for our MPS purpose; a typical sampled input contains $\eta N$ surviving photons on average, so the simulation cost remains exponential in $\eta N$ for standard boson-sampling simulation methods. 
Thus, this approach gives an efficient regime only when $\eta N=O(\log N)$, which is more restrictive than the known $O(1/\sqrt{N})$ transmission scaling for lossy boson sampling~\cite{garcia2019simulating}.

We instead use the following decomposition:
\begin{align}
    \hat{\sigma}
    =
    \frac{1}{2}|\psi_+\rangle\langle\psi_+|
    +
    \frac{1}{2}|\psi_-\rangle\langle\psi_-|,
    \label{eq: decomposition of spbs}
\end{align}
where
$|\psi_\pm\rangle
\equiv
\sqrt{1-\eta}|0\rangle
\pm
\sqrt{\eta}|1\rangle.$
With this decomposition, the total lossy input state becomes
\begin{align}
    \hat{\rho}_{\rm in}
    =
    \frac{1}{2^N}
    \sum_{s_1,\ldots,s_N\in\{-,+\}^N}
    |\psi_{\rm in}(s_1,\ldots,s_N)\rangle
    \langle\psi_{\rm in}(s_1,\ldots,s_N)|,
\end{align}
where $|\psi_{\rm in}(s_1,\ldots,s_N)\rangle\equiv\left(\bigotimes_{j=1}^N |\psi_{s_j}\rangle\right)\otimes|0\rangle^{\otimes(M-N)}.$
Thus, weak sampling from $\hat{\rho}_{\rm in}$ is equivalent to drawing a sign string $(s_1,\ldots,s_N)$ uniformly at random and then simulating the corresponding pure-state input.

The intuition behind the decomposition in Eq.~\eqref{eq: decomposition of spbs} is that the leading single-photon amplitude in each branch is mostly a coherent displacement, which does not contribute to entanglement across an output-mode bipartition. 
Specifically, for the $+$ branch, define the centered version of $|\psi_+\rangle$ by applying an appropriate displacement as
\begin{align}
    |\psi_{\rm c}\rangle
    \equiv
    \hat D\left(-\sqrt{\eta(1-\eta)}\right)|\psi_+\rangle,
\end{align}
satisfying
\begin{align}
    \langle\psi_{\rm c}|\hat a|\psi_{\rm c}\rangle
    =
    0,
    \qquad
    \langle\psi_{\rm c}|\hat a^\dagger\hat a|\psi_{\rm c}\rangle
    =
    \eta^2 .
\end{align}
Under a passive interferometer, the product displacement is mapped to an output displacement that factorizes across any output-mode bipartition and therefore does not change the Schmidt spectrum. 
Consequently, the entanglement-relevant centered part of each sampled branch has an effective photon number $O(\eta^2)$ per input mode. 
This reduction is the main reason why the decomposition in Eq.~\eqref{eq: decomposition of spbs} is useful for an MPS simulation. Similarly, one can easily check that in the multiphoton Fock and cat-state cases (see below), the effective remaining photon number of each branch is $O(\eta^2)$.

Finally, note that the two branches differ only by a phase shift:
$|\psi_-\rangle
=
e^{i\pi\hat n}
|\psi_+\rangle.$
Since $e^{i\pi\hat n}$ is a passive single-mode operation, choosing different signs in the sampled input only adds an input layer of phase shifters. This layer can be absorbed into the passive interferometer. Because the entropy bound below is uniform over arbitrary passive linear-optical unitaries, all sign branches obey the same bound. Therefore, for the entropy analysis, it is sufficient to consider the $+$ branch,
\begin{align}
    |\psi_+\rangle^{\otimes N}
    \otimes
    |0\rangle^{\otimes(M-N)} .
\end{align}

\begin{figure}[t]
    \centering
    \includegraphics[width=\columnwidth]{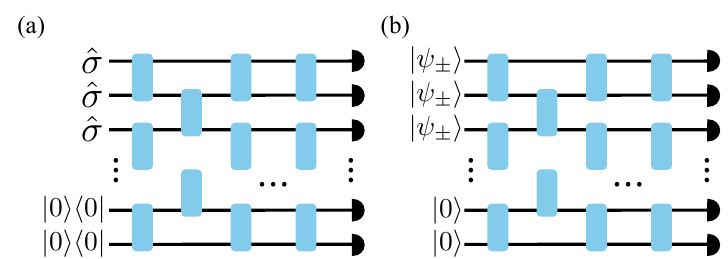}
    \caption{Simulation of lossy boson sampling using pure-state decomposition. (a) The Fock input $|1\rangle^{\otimes N}|0\rangle^{\otimes (M-N)}$ goes through the loss channel, yielding the product mixed state $\hat{\sigma}^{\otimes N}\otimes|0\rangle\langle 0|^{\otimes (M-N)}$, where $\hat{\sigma}$ is the lossy single-photon state. (b) For sampling, each $\hat{\sigma}$ is decomposed into $|\psi_+\rangle$ and $|\psi_-\rangle$, drawn with equal probability. }
    \label{fig3:BS_sampling}
\end{figure}

\subsection{R\'enyi entanglement entropy of the output state}
\label{subsection: ere of the output state}

We now bound the entanglement generated by a sampled pure branch under a passive interferometer. 
As explained above, it is sufficient to analyze the $+$ branch,
\begin{align}
    |\psi_{\rm in}\rangle
    =
    |\psi_+\rangle^{\otimes N}
    \otimes
    |0\rangle^{\otimes(M-N)} .
\end{align}
After the passive interferometer $\hat U$, the input state transforms to
\begin{align}
|\psi_{\rm out}\rangle
&=
\left[
\prod_{j=1}^{N}
\left(
\sqrt{1-\eta}
+
\sqrt{\eta}
\sum_{k=1}^{M}
U_{jk}^*
\hat a_k^\dagger
\right)
\right]
|0\rangle^{\otimes M}.
\label{eq:output of spbs}
\end{align}
Here, the operators $\hat a_k^\dagger$ denote the physical output-mode creation operators.
The entanglement is evaluated with respect to a bipartition $A:B$ of these physical output modes. 
The proof of the bound, given in Appendix~\ref{appendix: ere in spbs}, uses the following ingredients. 
First, we remove the coherent displacement of each sampled branch; after the interferometer, this displacement factorizes across any output-mode bipartition and therefore does not change the Schmidt spectrum. 
Second, for each input mode $j$, the transformed creation operator is split exactly according to the spatial bipartition,
\begin{align}
\sum_{k=1}^{M}U_{jk}^*\hat a_k^\dagger
=
\sum_{k\in A}U_{jk}^*\hat a_k^\dagger
+
\sum_{k\in B}U_{jk}^*\hat a_k^\dagger .
\end{align}
Finally, a Schmidt-rank counting argument for each Fock component, combined with Schatten-norm interpolation, yields a uniform bound over all passive interferometers and all spatial bipartitions of the output modes.

\begin{theorem}
\label{theorem:bounding ere}
Fix $1/2<\alpha<1$. For any number $N$ of single-photon input modes, any total number $M$ of modes, any passive linear-optical unitary $\hat U$, and any output-mode bipartition $A:B$, the sampled output branch in Eq.~\eqref{eq:output of spbs} satisfies
\begin{align}
S_\alpha(\hat{\rho}_A)
=
O\left(N\eta^{2\alpha}
\right),
\end{align}
where
$\hat{\rho}_A=\Tr_B\left(|\psi_{\rm out}\rangle\langle\psi_{\rm out}|\right).$
The implicit constant depends only on $\alpha$ and is uniform in $N$, $M$, $\hat U$, and the bipartition.
\end{theorem}

By the MPS approximability criterion reviewed in Sec.~\ref{sec:MPS}, it is sufficient that the R\'enyi entanglement entropy scales logarithmically across every MPS cut. 
Theorem~\ref{theorem:bounding ere} shows that this holds whenever, for fixed $1/2<\alpha<1$,
\begin{align}
    \eta
    =
    O\left[
    \left(
    \frac{\log N}{N}
    \right)^{1/(2\alpha)}
    \right].
\end{align}
Indeed, in this regime $N\eta^{2\alpha}=O(\log N)$, and hence the sampled pure branches admit polynomial-bond-dimension MPS approximations. 
Because the bound is uniform over all passive linear-optical unitaries and all output-mode bipartitions, it also controls the entanglement encountered at intermediate stages of any passive linear-optical circuit implementation. 
Thus, the above transmission scaling gives a sufficient regime for efficient MPS weak simulation of lossy boson sampling. 
Taking $\alpha$ arbitrarily close to $1$ approaches the familiar $1/\sqrt{N}$ transmission scaling, up to logarithmic factors.

We next show that the above MPS-approximability bound is essentially tight around the $1/\sqrt{N}$ scale.
Assume $M\geq 2N$, and consider the passive interferometer that applies a $50:50$ beam splitter between modes $j$ and $N+j$ for each $1\leq j\leq N$:
\begin{align}
    \hat a_j^\dagger
    \mapsto
    \frac{\hat a_j^\dagger+\hat a_{N+j}^\dagger}{\sqrt{2}} .
\end{align}
For the bipartition $A=\{1,\ldots,N\}$ and $B=\{N+1,\ldots,M\}$, the $+$ branch becomes a product of identical two-mode states,
\begin{align}
|\psi_{\rm out}\rangle =\bigotimes_{j=1}^{N}
\left(
\sqrt{1-\eta}|00\rangle_{j,N+j}
+
\sqrt{\frac{\eta}{2}}|10\rangle_{j,N+j}
\right.\nonumber\\
\left.+\sqrt{\frac{\eta}{2}}|01\rangle_{j,N+j}
\right)
\otimes
|0\rangle^{\otimes (M-2N)}.
\end{align}
For each pair, the reduced state on subsystem $A$ has eigenvalues
$\lambda_\pm=(1\pm\sqrt{1-\eta^2})/2.$
Hence, for $\alpha>1$,
\begin{align}
    S_\alpha(\hat\rho_A)
    =
    \frac{N}{1-\alpha}
    \log\left(
    \lambda_+^\alpha+\lambda_-^\alpha
    \right)
    \geq
    c_\alpha N\eta^2,
\end{align}
where $c_\alpha>0$ is a constant depending only on $\alpha$. The last inequality follows from
$\lambda_-\geq \eta^2/4$ and
$\lambda_+^\alpha+\lambda_-^\alpha\leq 1-\lambda_-(1-2^{1-\alpha})$.
Therefore, if
\begin{align}
    \eta
    =
    \Omega\left(N^{-1/2+\kappa/2}\right)
\end{align}
for some constant $\kappa>0$, then
\begin{align}
    S_\alpha(\hat\rho_A)
    =
    \Omega(N^\kappa).
\end{align}
By the complementary MPS criterion reviewed in Sec.~\ref{sec:MPS}, this state cannot be approximated to fixed accuracy with polynomial bond dimension across this cut. 
Together with Theorem~\ref{theorem:bounding ere}, this identifies the scale $\eta=\Theta(1/\sqrt{N})$, up to logarithmic factors, as the transition captured by our MPS analysis.

We now compare our result with previous classical algorithms for lossy boson sampling. 
Refs.~\cite{garcia2019simulating,oszmaniec2018classical} also obtained classical simulability around the transmission scaling $\eta=O(1/\sqrt{N})$, by approximating the lossy input state with a classically easy state, such as a thermal or separable state. 
These fixed-state approximations provide important asymptotic benchmarks. 
However, once the physical parameters are fixed, the approximating state is fixed as well. 
Thus, outside the sufficiently lossy regime, the simulation accuracy cannot be systematically improved by allocating more computational resources.

Our MPS approach reaches essentially the same asymptotic transmission scaling, up to logarithmic factors, but differs in its error--cost tradeoff. 
The algorithm remains well defined for all values of $\eta$: increasing the bond dimension improves the approximation, although the required bond dimension may become exponential outside the efficiently simulable regime. 
Thus, the main advantage of the present approach is not a better asymptotic threshold, but a tunable simulation framework that can be assessed quantitatively through the required bond dimension.

It is also useful to compare our approach with MPO-based simulations of lossy boson sampling~\cite{oh2021classical}. 
Both MPS and MPO methods have a tunable accuracy through the retained bond dimension. 
However, MPO simulations represent mixed states directly, which increases the local dimension and can also encode classical correlations in the bond dimension. 
For a circuit of depth $D$ with local cutoff $d$, the leading update costs scale as
$O(MDd^4\chi^3)$ for MPS and $O(MDd^8\chi^3)$ for MPO, while the memory costs scale as $O(Md\chi^2)$ and $O(Md^2\chi^2)$, respectively. 
This difference can be substantial in boson sampling, where a conservative local cutoff is $d=N+1$. 
Therefore, for the same target truncation criterion, the pure-state MPS representation can require significantly smaller computational resources than an MPO representation.

\subsection{Numerical analysis of the computational cost}\label{section: numerical computational cost}
Before presenting the resource estimates, we specify the truncation criterion used in the numerical analysis. 
We use the discarded Schmidt weight as a local MPS truncation criterion, as is standard in MPS simulations~\cite{schollwock2011density}. 
For the Schmidt decomposition across a cut $l$, the discarded weight after keeping the largest $\chi$ Schmidt coefficients is
\begin{align}
\epsilon_\chi^{(l)}
=
\sum_{\nu>\chi}
\left(s_\nu^{(l)}\right)^2 .
\end{align}
For a single normalized truncation, this gives a trace-distance error $\sqrt{\epsilon_\chi^{(l)}}$, and hence an upper bound $\sqrt{\epsilon_\chi^{(l)}}$ on the total-variation distance of any subsequent measurement outcome distribution~\cite{watrous2018theory}. 
In the estimates below, we set $\epsilon_\chi^{(l)}=0.01$ as a local threshold for estimating the bond dimension at the relevant bipartition. 
This should not be interpreted as a certified end-to-end sampling error: in a full MPS time-evolution simulation, local truncation errors must be accumulated over all relevant bonds and circuit layers.

Using this local criterion, we estimate the bond dimension and memory required for the sampled pure-state branches of lossy boson sampling as functions of the number of input photons $N$ and the transmission rate $\eta$. 
We focus on memory instead of wall-clock time because runtime depends strongly on implementation details, whereas the memory requirement follows directly from the estimated bond dimension. 
The resulting estimates should be understood as finite-size resource indicators for the proposed MPS simulation, rather than as rigorous global error guarantees for a complete implementation.

For the finite-size resource estimates, we do not perform a full
MPS time-evolution simulation, because obtaining the relevant Schmidt
spectra in that way would require carrying out the full simulation
whose cost we aim to estimate.
Instead, we use an approximate canonical-mode model to estimate the
Schmidt spectrum across a representative balanced bipartition.
This approximation is used only for the numerical resource analysis
and plays no role in the analytical bound of
Theorem~\ref{theorem:bounding ere}.

For the balanced bipartition of the physical output modes,
$A=\{1,\ldots,\lfloor M/2\rfloor\}$ and
$B=\{\lfloor M/2\rfloor+1,\ldots,M\}$, we write
\begin{align}
    \sum_{k=1}^M U_{jk}^*\hat a_k^\dagger
    =
    \cos\theta_j \hat B_{A,j}^\dagger
    +
    \sin\theta_j \hat B_{B,j}^\dagger,
\end{align}
where $\cos^2\theta_j \equiv \sum_{k\leq \lfloor M/2\rfloor}|U_{jk}|^2, \sin^2\theta_j\equiv\sum_{k> \lfloor M/2\rfloor}|U_{jk}|^2.$
Here, $\hat B_{A,j}^\dagger$ and $\hat B_{B,j}^\dagger$ act the two disjoint physical subsystems.

In the approximate canonical-mode model, we treat the projected modes
as independent canonical modes within each subsystem: $[\hat B_{A,j},\hat B_{A,k}^\dagger] \approx \delta_{jk}, [\hat B_{B,j},\hat B_{B,k}^\dagger]\approx \delta_{jk},[\hat B_{A,j},\hat B_{B,k}^\dagger]=0.$
Under this approximation, the sampled output branch takes the product form
\begin{align}
&|\psi_{\rm out}\rangle  \nonumber \\
&\approx
\bigotimes_{j=1}^N
\left[
\sqrt{1-\eta}|00\rangle_j
+
\sqrt{\eta}\cos\theta_j|10\rangle_j
+
\sqrt{\eta}\sin\theta_j|01\rangle_j
\right].
\end{align}
The corresponding reduced density matrix also factorizes, allowing us
to compute the approximate Schmidt spectrum directly.
We estimate the required bond dimension as the smallest $\chi$ for
which the retained eigenvalues have cumulative weight at least $0.99$,
equivalently corresponding to the local discarded-weight threshold
$\epsilon_\chi^{(l)}=0.01$.
The detailed procedure is given in
Appendix~\ref{appendix: subsection bond dimension analysis algorithm}.

We validate this approximation against exact bond-dimension calculations for Haar-random interferometers with $N\leq14$ in Appendix~\ref{appendix: subsection comparison between exact and est case}.
Within the tested regime, the approximate and exact estimates show
consistent behavior, with the approximate estimate being slightly larger for the Haar-random instances considered.
We therefore use the approximate canonical-mode model as a finite-size resource proxy.
Its apparent conservativeness is established only for the tested Haar-random instances, and we do not claim that this ordering holds universally or for the experimentally motivated architecture.

These estimates characterize only the resource requirements of the
present MPS method.
They should not be interpreted as a boundary for classical
simulability by all algorithms; in some regimes, methods such as the
Clifford--Clifford algorithm may be more efficient
~\cite{clifford2018classical,clifford2020faster}.

\begin{figure*}[t]
    \includegraphics[width=0.75\linewidth]{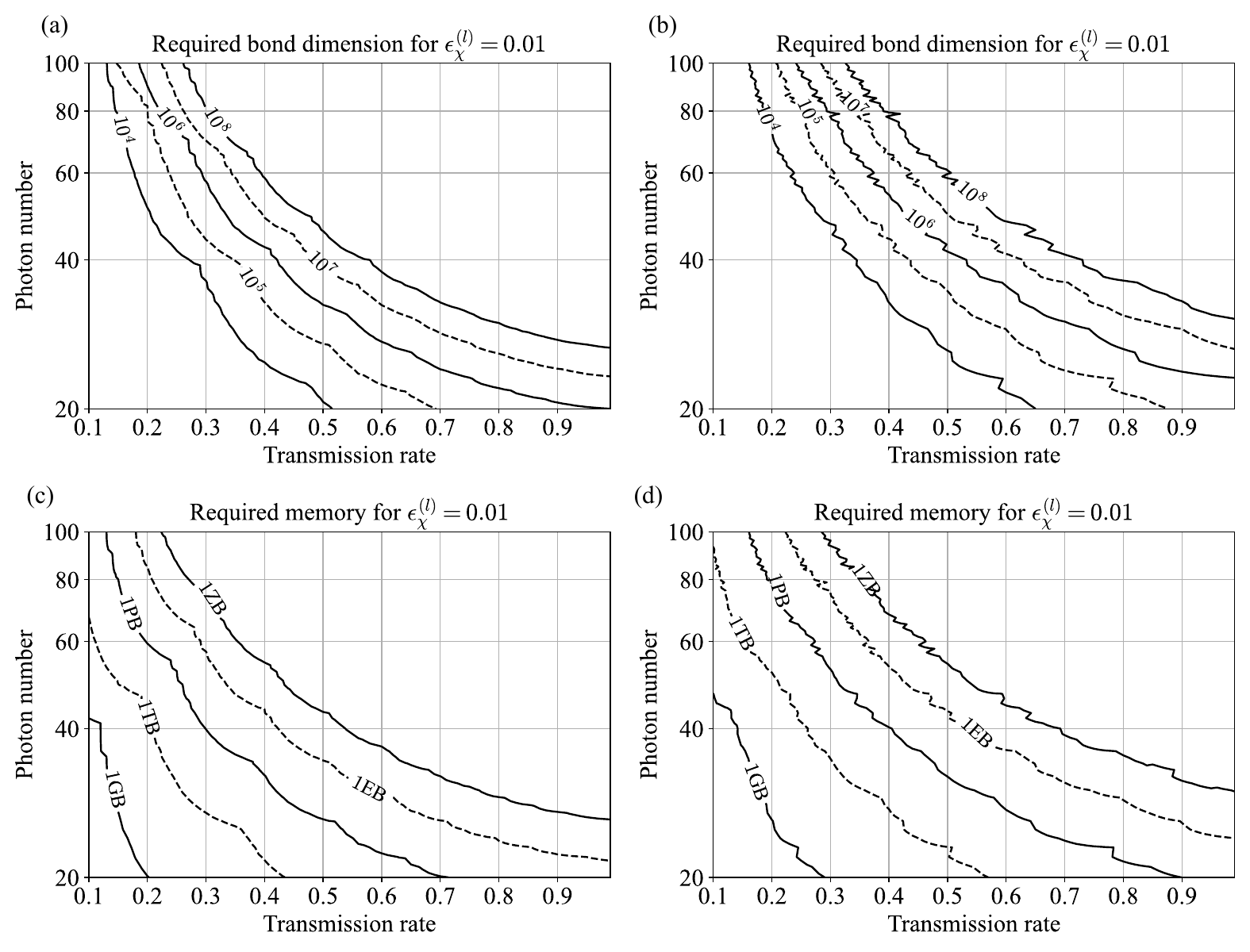}
 
    \caption{
    Required bond dimension and memory estimate for single-photon lossy boson sampling with local discarded-weight threshold $\epsilon_\chi^{(l)}=0.01$. Panels (a)--(b) show the estimated bond dimension for two types of interferometer: (a) Haar-random interferometers, and (b) experimentally motivated interferometers based on the architecture of Ref.~\cite{zhong2020quantum}. Panels (c)--(d) show the corresponding memory estimates. }
    \label{fig:numerical_analysis}
\end{figure*}

We present the numerical results in Fig.~\ref{fig:numerical_analysis}, comparing two interferometer families to illustrate how the required bond dimension depends on the circuit structure.
Panel~(a) uses Haar-random interferometers with $M=N^2$.
Panel~(b) uses representative unitaries constructed from the interferometer architecture of Ref.~\cite{zhong2020quantum}, although that experiment implements GBS.
Because this architecture permits only mode numbers that are twice a perfect square, for each $N$ we choose the allowed mode number closest to $N^2$.
We distribute the $N$ input photons as evenly as possible between the two halves of the interferometer and place them near the centers of the corresponding mode blocks, which produces relatively large entanglement within this architecture.

Figure~\ref{fig:numerical_analysis} shows that the estimated resource requirement of the present MPS method depends strongly on both the transmission rate and the interferometer structure.
Photon loss substantially reduces the required bond dimension, while the experimentally motivated architecture in panel~(b) generally requires fewer resources than the Haar-random
interferometers in panel~(a). 
For example, at $N=40$ and $\eta=0.5$, the estimated bond dimension decreases from approximately $10^7$ to below $10^6$, corresponding to factors of approximately $10^2$ and $10^3$ in
the leading memory and tensor-update cost estimates, respectively.

According to the required bond dimension, the corresponding MPS tensor-storage requirement can be estimated as shown in Fig.~\ref{fig:numerical_analysis}(c)--(d)~\cite{oh2024classical}.
For the single-photon boson-sampling instances considered here, we use the local Hilbert-space dimension $d=N+1$.
Assuming single-precision complex storage, corresponding to $8$ bytes per complex number, the memory requirement is estimated as
\begin{align}
&(\text{8 bytes})
\times
(\text{bond dimension})^2
\times
(\text{number of modes})
\nonumber\\
&\qquad\times
(\text{local Hilbert-space dimension}).
\end{align}

To provide concrete reference points for the memory-based feasibility of the present direct MPS method, we identify representative instances with terabyte-, petabyte-, and exabyte-scale storage requirements in Fig.~\ref{fig:numerical_analysis}(c). The points $(N,\eta)=(40,0.21)$ and $(30,0.26)$ require approximately $1\,\rm{TB}$ of tensor storage. We therefore regard these instances as memory-feasible on a sufficiently equipped computing server, although this estimate alone does not guarantee a practical runtime.

The points $(N,\eta)=(40,0.30)$ and $(60,0.20)$ require storage on the order of $1\,\rm{PB}$. Even taking $9.2\,\rm{PB}$ of aggregate supercomputer memory as an optimistic reference, these instances lie near the practical boundary of the present method, because storing the MPS tensors is only the baseline requirement: singular-value decompositions, tensor updates, and distributed-memory communication require substantial additional resources.

By contrast, the points $(N,\eta)=(60,0.32)$ and $(40,0.43)$
require storage on the order of $1\,\rm{EB}$. These instances are clearly infeasible for a direct MPS implementation from storage requirements alone.

For the experimentally motivated interferometer in Fig.~\ref{fig:numerical_analysis}(d), the estimated memory requirements are substantially smaller at the same parameter points. At $(N,\eta)=(40,0.21)$, the storage requirement
decreases from approximately $1\,\rm{TB}$ in panel (c) to $10\,\rm{GB}$, in panel
(d). Similarly, panel (d) gives approximately $8\,\mathrm{TB}$ at
$(N,\eta)=(40,0.30)$, $3.6\,\mathrm{PB}$ at $(40,0.43)$, and
$37\,\rm{PB}$ at $(60,0.32)$. Compared with the Haar-random interferometers in panel (c), the experimentally motivated architecture therefore reduces the estimated memory requirement by roughly one to two orders of magnitude at these representative points, corresponding to an approximately one-order-of-magnitude reduction in the required bond dimension. Nevertheless, the largest instances remain impractical for a direct MPS implementation despite this substantial reduction.

\subsection{Multiphoton Fock boson sampling}
\label{sec:fock}

We next consider multiphoton Fock-state inputs. 
This provides the simplest extension of the single-photon boson sampling and illustrates that the decomposition-based MPS strategy is not restricted to the latter. 
Suppose that each of the first $N$ modes is prepared in an $\ell$-photon Fock state, with fixed $\ell$, while the remaining $M-N$ modes are initialized in the vacuum. 
The interferometer and the final photon-number-resolving measurement are the same as in the single-photon boson-sampling setup.

After commuting the uniform loss to the input, the lossy input state is
$\hat{\rho}_{\rm in}
=
\hat{\sigma}^{\otimes N}
\otimes
|0\rangle\langle 0|^{\otimes(M-N)},$
where the lossy $\ell$-photon Fock state is
$\hat{\sigma}
=
\sum_{k=0}^\ell
\binom{\ell}{k}
(1-\eta)^{\ell-k}\eta^k
|k\rangle\langle k|.$
To apply the pure-state MPS strategy, we use the phase-averaged decomposition
\begin{align}
|\psi(\phi)\rangle
=
\sum_{k=0}^\ell
\sqrt{
\binom{\ell}{k}
\eta^k(1-\eta)^{\ell-k}
}
e^{ik\phi}
|k\rangle,
\end{align}
with $\phi$ sampled uniformly from $[0,2\pi)$. 
Averaging over the phase gives 
$\frac{1}{2\pi}\int_0^{2\pi}d\phi\,|\psi(\phi)\rangle\langle\psi(\phi)|=\hat{\sigma}$. 
Thus, a sampled pure input branch has the form
\begin{align}
\label{eq: fock sampled input}
    \left[
    \bigotimes_{j=1}^N
    |\psi(\phi_j)\rangle
    \right]
    \otimes
    |0\rangle^{\otimes(M-N)}.
\end{align}
Since $|\psi(\phi_j)\rangle=e^{i\phi_j \hat a_j^\dagger\hat a_j}|\psi(0)\rangle$, the sampled phases correspond to an input layer of phase shifters. 
This layer can be absorbed into the passive interferometer, and the entanglement bound is uniform over passive interferometers. 
Therefore, for the entropy analysis, it suffices to consider
$|\psi_{\rm in}\rangle
=
|\psi(0)\rangle^{\otimes N}
\otimes
|0\rangle^{\otimes(M-N)}.$
For an arbitrary spatial bipartition $A:B$ of the physical output modes, define the reduced state of the sampled output branch as $\hat{\rho}_A\equiv \Tr_B(\hat{U}|\psi_{\text{in}}\rangle\langle \psi_{\text{in}}|\hat{U}^\dagger)$.
For fixed $\ell$ and fixed $1/2<\alpha<1$, the same argument as in Theorem~\ref{theorem:bounding ere} then yields
\begin{align}
    S_\alpha\!\left(
    \hat{\rho}_A
    \right)=O(N\eta^{2\alpha}). 
\end{align}
Consequently, fixed-$\ell$ Fock-state boson sampling admits efficient MPS approximation in the same transmission regime,
\begin{align}
    \eta
    =
    O\left[
    \left(
    \frac{\log N}{N}
    \right)^{1/(2\alpha)}
    \right].
\end{align}
The input-dependent coefficient bound needed for this extension is given in Appendix~\ref{appendix: ere in fock}.

\begin{figure}[t]
    \centering
    \includegraphics[width=0.9\columnwidth]{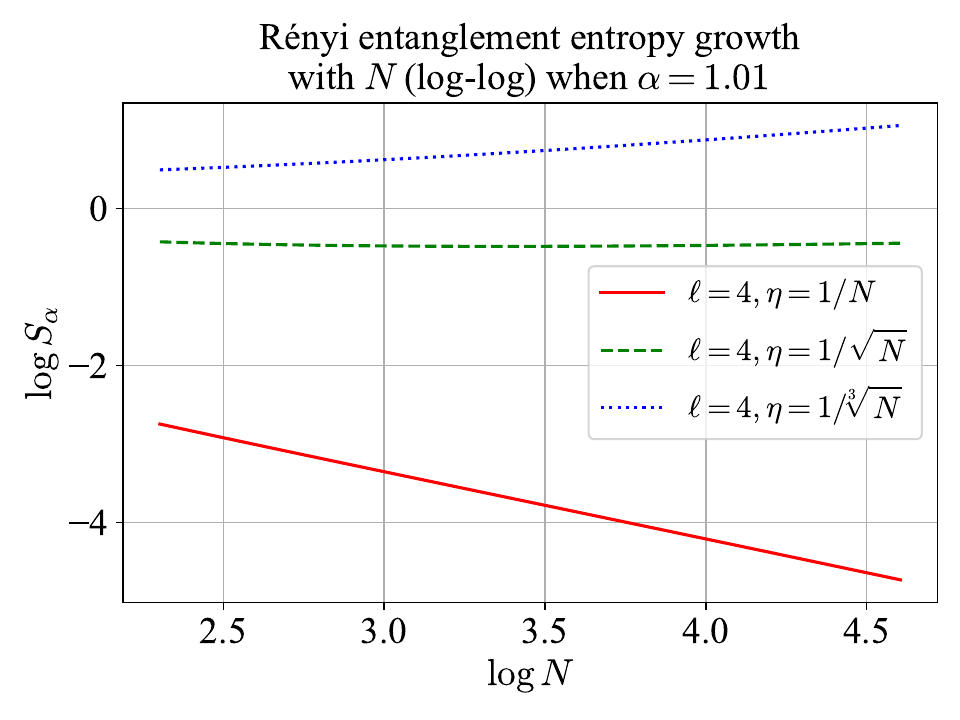}
    \caption{R\'enyi entanglement entropy as a function of the number $N$ of input modes, for fixed photon number $\ell$ per input mode and different transmission rates $\eta$. The $\alpha>1$ scaling shows that the $\eta=N^{-1/3}$ regime is inapproximable by polynomial bond dimension. }
    \label{fig:fock}
\end{figure}

To probe MPS approximability beyond the sufficient regime identified above, we use the same equal-splitting interferometer as in the single-photon case and numerically evaluate the R\'enyi entropy for $\alpha>1$; see Fig.~\ref{fig:fock}.
The interferometer acts as $\hat a_j^\dagger\mapsto\frac{\hat a_j^\dagger+\hat a_{N+j}^\dagger}{\sqrt{2}}$ for $1\leq j\leq N$.
For the example $\eta=N^{-1/3}$, the log--log plot shows approximately linear growth of the $\alpha>1$ R\'enyi entropy with $N$, indicating algebraic scaling $S_\alpha=\Omega(N^\kappa)$ for some $\kappa>0$. 
By the complementary MPS criterion in Sec.~\ref{sec:MPS}, this implies that the corresponding output state cannot be approximated to fixed accuracy with polynomial bond dimension across this cut. 
Thus, fixed-$\ell$ Fock-state inputs exhibit the same qualitative MPS-approximability transition as the single-photon case.

\subsection{Cat-state boson sampling}
We also consider lossy boson sampling with even and odd cat-state inputs of fixed amplitude $\gamma$, defined as
\begin{align}
    |\text{cat}_\pm\rangle &\equiv \frac{1}{\sqrt{2 \left( 1 \pm e^{-2|\gamma|^2} \right)}} (|\gamma\rangle \pm |-\gamma\rangle).
\end{align}
Without loss of generality, we take $\gamma>0$ to be real, since its phase can be absorbed into input phase shifters and hence into the passive interferometer.
Note that as the amplitude of the cat state $\gamma\rightarrow 0$, the odd~(even) cat state approaches $|1\rangle$~($|0\rangle$), thereby the odd-cat input recovers a single-photon input. Moreover, cat state sampling is also known to be hard to classically simulate~\cite{rohde2013sampling}. While we focus on odd cat-state input, an analogous analysis can be applied to even cat-state input and shows a similar result. 
The interferometer and measurement are the same as in the single-photon case. 
Uniform photon loss can be commuted to the input of the circuit as in Fig.~\ref{fig1:setup_figure}(a)--(b), and the cat state input goes through the total photon loss channel and transforms to the following state, 
\begin{align}
\hat{\rho}_{\text{in}}=\hat{\sigma}^{\otimes N}\otimes |0\rangle\langle 0|^{\otimes (M-N)}, 
\end{align}
where the noisy cat state becomes
\begin{align}
    \hat{\sigma}&\equiv\frac{1}{2(1-e^{-2|\gamma|^2})}\Bigl[|\gamma\sqrt{\eta}\rangle\langle\gamma\sqrt{\eta}|+|-\gamma\sqrt{\eta}\rangle\langle -\gamma\sqrt{\eta}| \nonumber\\
    &-e^{-2|\gamma|^2(1-\eta)}(|\gamma\sqrt{\eta}\rangle\langle-\gamma\sqrt{\eta}|+|-\gamma\sqrt{\eta}\rangle \langle\gamma\sqrt{\eta}|)\Bigr].
\end{align}
We use the following state decomposition of the lossy cat state:
\begin{align}\label{eq:lossy cat state decomposition}
\hat{\sigma}=\dfrac{1}{2}|\psi_+\rangle\langle\psi_+| + \dfrac{1}{2}|\psi_-\rangle\langle\psi_-|, 
\end{align}
where we define
\begin{align}\label{eq: sampled pure state input of cat state}
|\psi_\pm\rangle&\equiv \pm a_\pm|\gamma\sqrt{\eta}\rangle\mp a_\mp|-\gamma\sqrt{\eta}\rangle,\\
a_\pm\equiv &\dfrac{\sqrt{1+e^{-2|\gamma|^2(1-\eta)}}\pm\sqrt{1-e^{-2|\gamma|^2(1-\eta)}}}{\sqrt{4(1-e^{-2|\gamma|^2})}}.
\end{align}

Since $|\psi_-\rangle=e^{i\pi\hat n}|\psi_+\rangle$, the two sampled branches differ only by an input phase shift, which can be absorbed into the passive interferometer.
It therefore suffices to analyze
\begin{align}
    |\psi_+\rangle^{\otimes N}
    \otimes
    |0\rangle^{\otimes(M-N)}.
\end{align}
For any fixed $1/2<\alpha<1$, the same argument as in Theorem~\ref{theorem:bounding ere} gives
\begin{align}
    S_\alpha(\hat{\rho}_A)
    =
    O(N\eta^{2\alpha})
\end{align}
for every output-mode bipartition $A:B$.
Consequently, for fixed $\gamma$, lossy cat-state boson sampling admits efficient MPS approximation when
\begin{align}
    \eta
    =
    O\left[
    \left(
    \frac{\log N}{N}
    \right)^{1/(2\alpha)}
    \right].
\end{align}

\section{Noisy IQP sampling}
\label{sec:IQP}
\subsection{Decomposition of noisy input state}
We now apply the decomposition-based MPS strategy to noisy IQP sampling. We first consider dephasing noise and later extend the analysis to depolarizing noise. As described in Sec.~\ref{sec:IQPsetup}, the dephasing channel commutes with all diagonal gates in an IQP circuit. Hence, if each of the $d$ layers is followed by dephasing noise with rate $p$, all dephasing channels can be moved to the input and combined into a single dephasing channel $\mathcal{N}_{0,0,p_d}$, where
\begin{align}
p_d
=
\frac{1-(1-2p)^d}{2}.
\end{align}
The noisy circuit is therefore equivalent to this input noise channel followed by an ideal IQP circuit, as illustrated in Fig.~\ref{fig1:setup_figure}(c)--(d).

After this input noise, the initial state becomes $\hat{\rho}_{\rm in}=\hat{\tau}^{\otimes n},$ where
\begin{align}
\hat{\tau}
\equiv
\mathcal{N}_{0,0,p_d}(|+\rangle\langle +|)
=
(1-p_d)|+\rangle\langle +|
+
p_d|-\rangle\langle -|.
\end{align}

As in the lossy boson-sampling case, the decomposition of the mixed input is crucial. 
The diagonal decomposition of $\hat{\tau}$, $\hat{\tau}=(1-p_d)|+\rangle\langle +|+p_d|-\rangle\langle -|,$ is natural but not useful for our MPS purpose. 
Indeed, $|-\rangle=\hat Z|+\rangle$, so sampling this decomposition only inserts input $Z$ gates, which can be absorbed into the diagonal IQP circuit. 
Thus, each sampled branch is essentially another noiseless IQP circuit with $|+\rangle^{\otimes n}$ input, and the noise does not reduce the entanglement relevant to the MPS simulation.

Instead, we use the decomposition
\begin{align}
\hat{\tau}=\frac{1}{2}|\psi_+\rangle\langle \psi_+|+\frac{1}{2}|\psi_-\rangle\langle \psi_-|,
\end{align}
where
$|\psi_\pm\rangle\equiv \sqrt{1-p_d}|+\rangle\pm\sqrt{p_d}|-\rangle
= w_\pm|0\rangle+w_\mp|1\rangle$, 
with $w_\pm\equiv(\sqrt{1-p_d}\pm\sqrt{p_d})/\sqrt{2}$. Thus, the total noisy input state is written as
\begin{align}
    \hat{\rho}_{\text{in}}=\dfrac{1}{2^n}\sum\limits_{s_1,\dots,s_n\in\{-,+\}^n}|\psi_{\text{in}}(s_1,\dots,s_n)\rangle\langle\psi_{\text{in}}(s_1,\dots,s_n)|,
\end{align}
where $|\psi_\text{in}(s_1,\dots,s_n)\rangle\equiv\bigotimes_{j=1}^n |\psi_{s_j}\rangle.$
Therefore, weak sampling from $\hat{\rho}_{\rm in}$ can be performed by drawing a sign string uniformly at random and then simulating the corresponding pure-state IQP circuit.

For the entanglement analysis, it is enough to consider the all-$+$ branch. 
Indeed, $|\psi_-\rangle=\hat X|\psi_+\rangle$, so changing the sampled signs only applies input bit flips. 
These bit flips merely permute the computational-basis amplitudes of the sampled branch and do not change the R\'enyi-entropy bound derived below. 
Thus, we take
\begin{align}
|\psi_{\rm in}\rangle
=
|\psi_+\rangle^{\otimes n}
=
\sum_{\vec{x}\in\{0,1\}^n}
w_+^{n-|\vec{x}|}
w_-^{|\vec{x}|}
|\vec{x}\rangle .
\end{align}

\begin{figure}[t]
    \centering
    \includegraphics[width=\columnwidth]{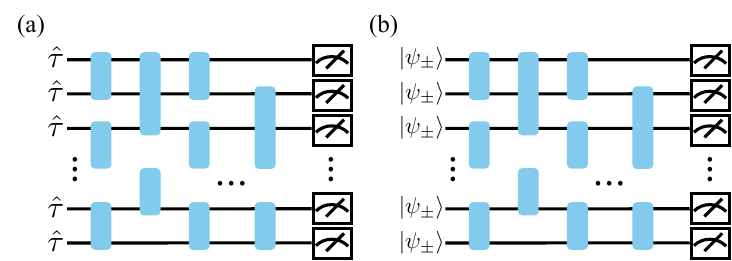}
    \caption{Pure-state sampling from mixed inputs in an IQP circuit. (a) The input $|+\rangle$ goes through the noise channel and becomes a mixed state $\hat{\tau}$. (b) We sample from an ensemble decomposition of the mixed state, drawing $|\psi_+\rangle$ or $|\psi_-\rangle$ with equal probability $1/2$. }
    \label{fig:IQP_sampling}
\end{figure}

\subsection{R\'enyi entanglement entropy of the output state}

We now bound the entanglement of the sampled branch after the ideal IQP circuit. 
Since an IQP circuit is diagonal in the computational basis, it maps each basis state as
$|\vec{x}\rangle\mapsto e^{if(\vec{x})}|\vec{x}\rangle$. 
Thus,
\begin{align}
|\psi_{\rm out}\rangle
=
\sum_{\vec{x}\in\{0,1\}^n}
e^{if(\vec{x})}
w_+^{n-|\vec{x}|}
w_-^{|\vec{x}|}
|\vec{x}\rangle .
\end{align}

For a bipartition $A:B$, let $\hat{\rho}_A=\Tr_B(|\psi_{\text{out}}\rangle\langle\psi_{\text{out}}|)$ be the reduced density matrix of the output state. 
Its spectrum generally depends on the IQP phase function $f$. 
To obtain a circuit-independent upper bound, we apply the completely dephasing channel $\Phi$ on subsystem $A$. 
Since $\Phi$ is unital, the eigenvalues of $\Phi(\hat{\rho}_A)$ are majorized by those of $\hat{\rho}_A$~\cite{watrous2018theory}. 
Because the R\'enyi entropy is Schur-concave for $\alpha>0$~\cite{marshall1979inequalities}, this gives $S_\alpha(\hat{\rho}_A)\leq S_\alpha(\Phi(\hat{\rho}_A))$.
The dephased state $\hat{\rho}_{A,\rm diag}\equiv\Phi(\hat{\rho}_A)$ is independent of the phase function $f$ and is a product state with single-qubit probabilities $w_+^2$ and $w_-^2$. Therefore,
\begin{align}
    S_\alpha(\hat{\rho}_A)
    \leq
    S_\alpha(\hat{\rho}_{A,\rm diag})
    =
    \frac{|A|}{1-\alpha}
    \log\left(w_+^{2\alpha}+w_-^{2\alpha}\right).
\end{align}
Applying the same argument after tracing out subsystem $A$, the bound is controlled by the smaller side of the bipartition. Thus, for any MPS cut $l$,
\begin{align}
    S_\alpha^l(|\psi_{\rm out}\rangle)
    &\leq
    \frac{n}{2(1-\alpha)}
    \log\left(w_+^{2\alpha}+w_-^{2\alpha}\right) \\
    &\leq
    \frac{n}{2(1-\alpha)}
    (1-2p)^{2\alpha d}.
\end{align}
Here, we used $w_+^{2\alpha}\leq 1$, $\log(1+x)\leq x$, and
$w_-
=
\frac{\sqrt{1-p_d}-\sqrt{p_d}}{\sqrt{2}}
\leq
1-2p_d
=
(1-2p)^d.$

By the MPS approximability criterion in Sec.~\ref{sec:MPS}, efficient MPS approximation is guaranteed when this bound scales as $O(\log n)$, namely when
\begin{align}
    n(1-2p)^{2\alpha d}
    =
    O(\log n).
\end{align}
Equivalently, up to logarithmic corrections and constants depending on $\alpha$, it is sufficient that
\begin{align}
    d
    =
    \Omega\left(
    \frac{\log n}{|\log(1-2p)|}
    \right).
\end{align}
For small $p$, this becomes $d=\Omega(p^{-1}\log n)$. Thus, once the depth is logarithmic in $n$ for fixed noise rate, the accumulated dephasing noise makes the sampled branches efficiently approximable by MPS. This scaling is consistent with recent results obtained by different methods~\cite{nelson2024polynomial}.

\begin{figure*}[t]
    \centering
    \includegraphics[width=\linewidth]{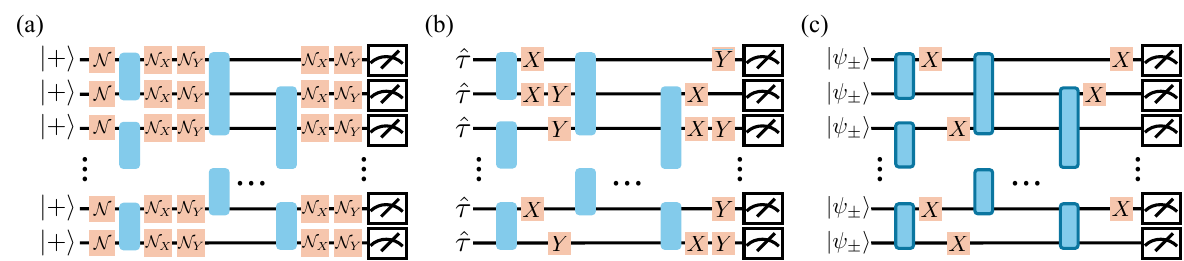}
    \caption{Simulation of depolarizing noise in an IQP circuit. (a) Pauli-$Z$ noise channels commute with all IQP gates and can therefore be moved to the input. (b) The noise channels are sampled as Pauli gates, with $X$ and $Y$ errors each occurring with probability $q$, while the input $|+\rangle$ passes through the Pauli $Z$ noise channel, resulting in the mixed state $\hat{\tau}$. (c) We sample from an ensemble decomposition of $\hat{\tau}$, drawing $|\psi_+\rangle$ or $|\psi_-\rangle$ each with equal probability $1/2$. Each $Y$ gate can be decomposed into $X$ and $Z$ gates. Here, because IQP gates are diagonal in the $Z$ basis, all $Z$ gates can be absorbed into the ideal IQP layer, as indicated by the shaded boundary. In contrast, $X$ gates do not commute with the IQP gates and therefore remain in the circuit.}
    \label{fig:IQP_sim_depol}
\end{figure*}

\subsection{Extension to depolarizing noise}
\label{sec:depol}
We now extend the dephasing analysis to depolarizing noise. Recall from Sec.~\ref{sec:IQPsetup} that a depolarizing channel with parameter $p$ can be decomposed into single-Pauli noise channels with rate $q=(1-\sqrt{1-2p})/2$. The Pauli-$Z$ part commutes with all diagonal IQP gates and can therefore be moved to the input, exactly as in the dephasing case. For a depth-$d$ circuit, this gives an accumulated input $Z$-noise rate $q_d=(1-(1-2q)^d)/2$. Applying the same input-state decomposition as before, we may take the sampled input branch to be
\begin{align}
|\psi_{\rm in}\rangle
=
|\psi_+\rangle^{\otimes n}
=
\sum_{\vec{x}\in\{0,1\}^n}
w_+^{n-|\vec{x}|}
w_-^{|\vec{x}|}
|\vec{x}\rangle,
\end{align}
where $w_\pm=(\sqrt{1-q_d}\pm\sqrt{q_d})/\sqrt{2}$. The remaining Pauli-$X$ and Pauli-$Y$ parts are treated as sampled local Pauli gates inside the circuit, as illustrated in Fig.~\ref{fig:IQP_sim_depol}.

After sampling the Pauli errors, every $Y$ gate can be written as $Y=-iZX$. 
The $Z$ parts can be absorbed into the diagonal IQP circuit, modifying only the phase function. 
The remaining $X$ gates act as computational-basis bit flips. 
Thus, each sampled Pauli-noise realization can be written as an ideal IQP circuit with a modified phase function, together with a bit-flip pattern $\vec u\in\{0,1\}^n$. 
Accordingly, each sampled pure-state branch has the form
\begin{align}
|\psi_{\rm out}\rangle
=
\sum_{\vec{x}\in\{0,1\}^n}
e^{if_{\vec u}(\vec{x})}
w_+^{n-|\vec{x}\oplus\vec u|}
w_-^{|\vec{x}\oplus\vec u|}
|\vec{x}\rangle .
\end{align}

For a bipartition $A:B$, write $\vec u_A$ for the restriction of the bit-flip pattern to subsystem $A$. 
After applying the same dephasing upper bound as before, the diagonal state on $A$ is
\begin{align}
    \hat{\rho}_{A,\rm diag}
    =
    \sum_{\vec{x}_A\in\{0,1\}^{|A|}}
    w_+^{2|A|-2|\vec{x}_A\oplus\vec u_A|}
    w_-^{2|\vec{x}_A\oplus\vec u_A|}
    |\vec{x}_A\rangle\langle\vec{x}_A| .
    \label{eq:xgate_iqp}
\end{align}
The bit flip $\vec u_A$ only permutes the diagonal entries. 
Therefore, $\hat{\rho}_{A,\rm diag}$ has the same eigenvalue spectrum as in the dephasing case, and the same R\'enyi-entropy bound applies. 
Consequently, depolarizing-noisy IQP sampling is efficiently approximable by MPS whenever
\begin{align}
    d
    =
    \Omega\left(
    \frac{\log n}{|\log(1-2q)|}
    \right),
\end{align}
where $q=(1-\sqrt{1-2p})/2$ is the single-Pauli noise rate in the depolarizing decomposition. 
For small $p$, this gives the scaling $d=\Omega(p^{-1}\log n)$, consistent with the critical-depth scaling in Ref.~\cite{nelson2024polynomial}.

\subsection{Numerical analysis of noisy IQP sampling}

\begin{figure*}[t]
    \centering
    \includegraphics[width=\linewidth]{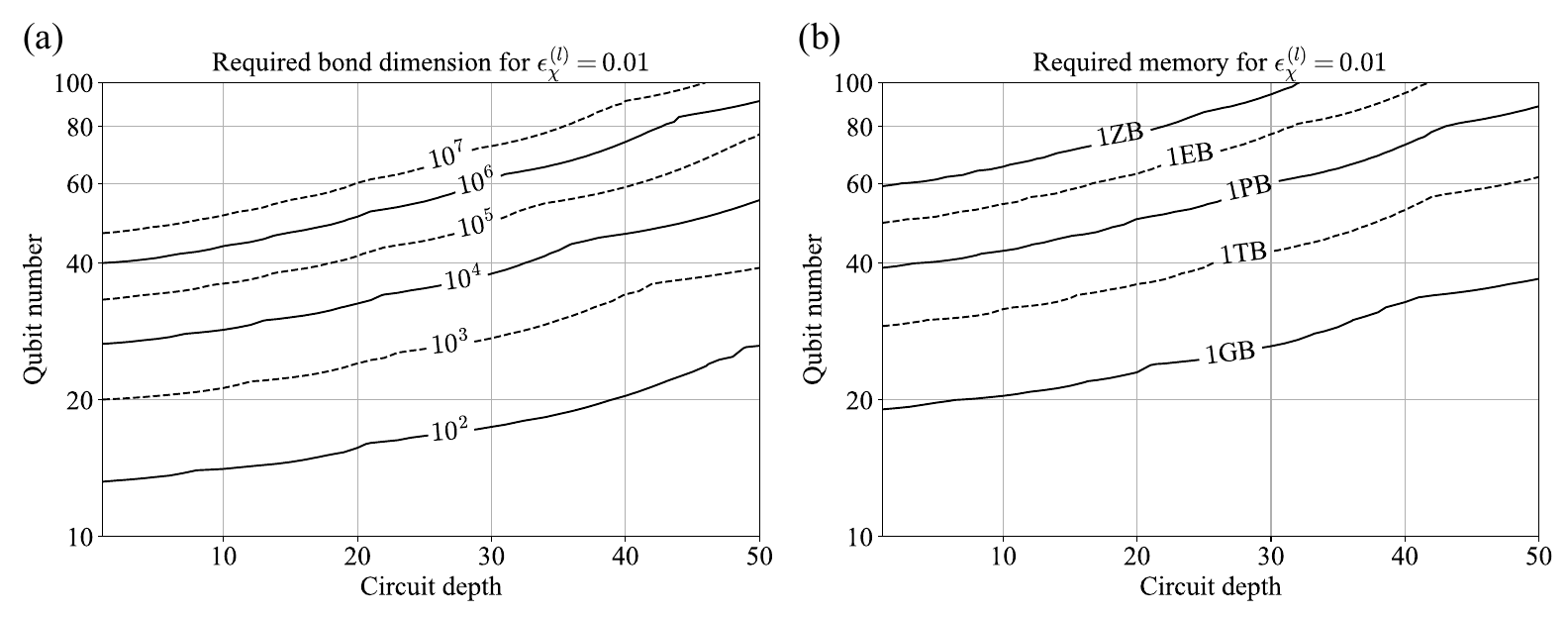}
    \caption{Estimated resources for dephasing-noisy IQP sampling with local discarded-weight threshold $\epsilon_\chi^{(l)}=0.01$. 
    (a) Estimated bond dimension and (b) corresponding memory estimate as functions of the qubit number and circuit depth, for dephasing noise rate $p=0.01$ per layer. 
    The estimates use the phase-independent diagonal upper-bound spectrum for the balanced bipartition $|A|=|B|=n/2$.}
    \label{fig:IQP_resource_estimation}
\end{figure*}

We now estimate the bond dimension and memory required for dephasing-noisy IQP sampling using the entropy bound derived above. 
As in the boson-sampling resource estimates, we use the local discarded-weight threshold $\epsilon_\chi^{(l)}=0.01$ to estimate the required bond dimension. 
For each pair of qubit number $n$ and circuit depth $d$, we use the diagonal upper-bound spectrum of $\hat{\rho}_{A,\rm diag}$ for the balanced cut $|A|=|B|=n/2$. 
Thus, the estimates in Fig.~\ref{fig:IQP_resource_estimation} should be interpreted as resource estimates for the phase-independent MPS upper bound, rather than as the cost of simulating a specific depth-$d$ IQP architecture.

Because the estimate is based on a phase-independent upper bound, it does not account for restrictions imposed by a particular finite-depth IQP architecture. 
In very shallow circuits, entanglement cannot spread across the system arbitrarily, and specialized classical simulation methods may be more efficient~\cite{markov2008simulating,bremner2017achieving,oh2022classical,maslov2024fast}. 
Therefore, the large bond dimensions shown at very small depths should be interpreted as an overestimate from our worst-case entropy bound, not as evidence that all such shallow IQP circuits are difficult for MPS methods. 
Away from this shallow-depth regime, increasing the depth also increases the accumulated dephasing noise, which suppresses the coefficient $w_-$ and reduces the estimated bond dimension.

It is also useful to compare this behavior with percolation-based simulability results for noisy IQP circuits. 
For example, Ref.~\cite{rajakumar2025polynomial} identifies a critical-depth regime beyond which noisy IQP circuits become classically simulable by their method. 
Our MPS estimate gives a complementary resource-based perspective: even before the asymptotic percolation threshold is reached, the required bond dimension can already be moderate for finite system sizes. 
For example, for $n=100$ qubits and local dimension $2$,
a bond dimension $\chi=10^4$ corresponds to approximately $(8~\mathrm{bytes})\times 100\times 2\times(10^4)^2\approx 160\,\mathrm{GB}$ under the same storage convention as in the boson-sampling analysis. 
Thus, parts of the parameter regime with bond dimensions of order $10^4$ remain plausible from a memory standpoint on large-memory classical hardware.
This does not by itself guarantee a practical runtime or a certified end-to-end sampling error, but it demonstrates that the finite-size resource estimate can be substantially more informative than the asymptotic threshold alone.

\section{Discussion and conclusion}
\label{sec:discussion}

In this work, we generalized the decomposition-based MPS strategy, originally developed for lossy GBS, to broader classes of sampling problems, including boson sampling and IQP sampling.
This generalization shows that the approach is not tied to Gaussian structure alone, and can be extended to other sampling models when suitable pure-state decompositions are available.

The proposed algorithm provides a tunable MPS-based simulation framework and can be more resource-efficient than mixed-state MPO approaches in the regimes considered here.
In addition to its efficiency, the method offers further advantages, such as tunable simulation accuracy that depends on the available computational resources. Notably, the approach remains valid across the entire range of imperfections, even in regimes where the simulation complexity increases from polynomial to exponential with respect to system parameters. 
Hence, we expect that the method can be used as a practical finite-size benchmark for quantum-advantage experiments in regimes where the required bond dimension remains manageable.

Furthermore, we provide finite-size resource estimates by translating the bond dimensions required under a target local discarded-weight criterion into approximate memory costs. These estimates are intended as memory-based feasibility indicators for the present direct MPS strategy, rather than as end-to-end runtime or global-error guarantees; the boson-sampling estimates additionally rely on the approximate canonical-mode model. In particular, we consider an experimentally motivated interferometer constructed from the architecture of Ref.~\cite{zhong2020quantum} to illustrate how the estimated resource requirements depend on circuit structure.

We remark that several extensions remain open. For lossy boson sampling, although our method covers single-photon, Fock-state, and cat-state inputs, whether it can be further generalized to arbitrary input states remains open. For IQP sampling, our analysis focuses on conventional diagonal IQP circuits~\cite{bremner2011classical}; it does not immediately extend to CNOT-augmented IQP circuits such as those considered in
Ref.~\cite{nelson2024polynomial}. More broadly, the present method is most directly applicable when the noise can be propagated to the input or converted into sampled local operations without leaving a circuit family whose MPS entanglement can be controlled. Extending the approach to more general noise channels, especially noncommuting noise that must be treated during the MPS evolution rather than solely at the input, is an important open question.

\section*{Funding}
S.P. and C.O. were supported by the National Research Foundation of Korea Grants (No. RS-2024-00431768 and No. RS-2025-00515456) funded by the Korean government (Ministry of Science and ICT (MSIT)) and the Institute of Information \& Communications Technology Planning \& Evaluation (IITP) Grants funded by the Korean government (MSIT) (No. RS-2024-00437284, No. IITP-2025-RS-2025-02283189 and No. IITP-2025-RS-2025-02263264). This work was supported by Global Partnership Program of Leading Universities in Quantum Science and Technology (RS-2025-08542968) through the National Research Foundation of Korea~(NRF) funded by the Korean government (Ministry of Science and ICT(MSIT)).

\section*{Data availability}
All data supporting the findings of this study are available in the Zenodo repository at:
\url{https://doi.org/10.5281/zenodo.21149576}.

\section*{Author contributions}
S.P. carried out the detailed calculations.
C.O. conceptualized the idea and supervised the overall project.
All authors contributed to writing the manuscript.

\section*{Competing interests}
The authors declare no competing interests.

\bibliography{reference.bib}

\onecolumngrid

\appendix

\section{Bound on R\'enyi entanglement entropy in single-photon boson sampling}\label{appendix: ere in spbs}

In this Appendix, we prove the R\'enyi entanglement entropy
bound stated in Sec.~\ref{subsection: ere of the output state} for sampled single-photon inputs.
The proof is uniform over the passive interferometer and the spatial bipartition of the physical output modes.

\begin{theorem-restate}[Uniform entanglement bound for sampled lossy inputs]

\label{theorem:single photon ere appendix}
Fix $1/2<\alpha<1$.
For any number $N$ of nonvacuum input modes, any total number $M$ of modes, any passive linear-optical unitary $\hat U$, and any spatial bipartition $A:B$ of the physical output modes, there exists a constant $C_\alpha<\infty$,
depending only on $\alpha$, such that the reduced state $\hat\rho_A$ of the sampled output branch satisfies 
\begin{align}
    \Tr(\hat{\rho}_A^\alpha)\leq (1+C_\alpha\eta^{2\alpha})^N.
\end{align}
Consequently, with $K_\alpha\equiv C_\alpha/(1-\alpha)$,
\begin{align}
    S_\alpha(\hat{\rho}_A)\leq \frac{N}{1-\alpha}\log (1+C_\alpha\eta^{2\alpha})\leq K_\alpha N\eta^{2\alpha}.
\end{align}

\end{theorem-restate}
For lossy fixed-input-photon-number Fock-state boson sampling and fixed-amplitude cat-state sampling, the same argument applies after replacing the sampled lossy single-photon branch with the corresponding sampled branch, namely, $|\psi(0)\rangle$ and $|\psi_+\rangle$, respectively. In these cases, the constants may depend on the fixed input parameters, such as the photon number $\ell$ or the cat-state amplitude $\gamma$. Thus, all three input families obey the same R\'enyi entanglement entropy scaling and hence yield the same efficiently simulable transmission-rate scaling under this MPS criterion.

We now prove the theorem. Throughout the proof, the bipartition $A:B$ is a spatial bipartition of the physical output modes. The delocalized, transformed modes are used only as an auxiliary orthonormal basis for expanding the evolved state. The proof does not assume that the projections of these modes onto $A$ and $B$ form independent canonical mode families.

The proof proceeds in four steps.
First, we remove the coherent displacement from each sampled input component and show that it becomes a local unitary across the output bipartition.
Second, we expand the centered output state in an orthonormal transformed-mode Fock basis and represent it by a coefficient matrix across the bipartition.
Third, for a finite Fock-space truncation, we combine a Schmidt-rank bound with Schatten-norm interpolation.
Finally, we pass to the full infinite expansion by proving convergence in the relevant Schatten class.

\subsection{Removing the coherent part by a local displacement}
We first remove the coherent displacement of each sampled input component and obtain the centered state $|\psi_{\rm c}\rangle$. This step is useful because the leading $O(\sqrt{\eta})$ contribution in $|\psi_+\rangle$ is a coherent displacement and therefore does not contribute to the entanglement across an output-mode bipartition.

Recall that
\begin{align}
    |\psi_+\rangle= \sqrt{1-\eta}|0\rangle+\sqrt{\eta}|1\rangle .
\end{align}

It suffices to analyze the all-$+$ branch.
Indeed, replacing any $|\psi_+\rangle$ by $|\psi_-\rangle$
amounts to an input phase shift, which can be absorbed into the passive interferometer.
Because the bound is uniform over passive interferometers, the same estimate holds for every sampled sign string.
Its displacement amplitude is
$\mu
\equiv
\langle\psi_+|\hat a|\psi_+\rangle
=
\sqrt{\eta(1-\eta)}.$
We define the centered single-mode state by
\begin{align}
    |\psi_{\rm c}\rangle
    \equiv
    \hat D(-\mu)|\psi_+\rangle
    =
    \sum_{m=0}^{\infty} c_m |m\rangle ,
\end{align}
where $\hat D(\mu)=\exp(\mu \hat a^\dagger-\mu \hat a)$ and $\mu$ is real. 
Equivalently, $|\psi_+\rangle=\hat D(\mu)|\psi_{\rm c}\rangle.$

For the $N$ non-vacuum input modes, this gives
\begin{align}
    |\psi_+\rangle^{\otimes N}\otimes |0\rangle^{\otimes(M-N)}
    =
    \hat D_{\rm in}(\mu)
    \left(
    |\psi_{\rm c}\rangle^{\otimes N}\otimes |0\rangle^{\otimes(M-N)}
    \right),
\end{align}
where $\hat D_{\rm in}(\mu) \equiv \bigotimes_{j=1}^{N}\hat D_j(\mu) \otimes \bigotimes_{j=N+1}^{M}\hat I_j.$
After the passive interferometer, the output state can be written as
\begin{align}
    \hat U
    \left(
    |\psi_+\rangle^{\otimes N}\otimes |0\rangle^{\otimes(M-N)}
    \right)
    =
    \hat D_{\rm out}(\mu')
    |\Phi\rangle , \qquad
    \text{where} \qquad
    |\Phi\rangle
    \equiv
    \hat U
    \left(
    |\psi_{\rm c}\rangle^{\otimes N}\otimes |0\rangle^{\otimes(M-N)}
    \right),
\end{align}
and $\hat D_{\rm out}(\mu')\equiv\hat U \hat D_{\rm in}(\mu)\hat U^\dagger$ is a multimode displacement on the physical output modes.

Now fix an arbitrary spatial bipartition on the output $A:B$ whose R\'enyi entanglement entropy bound will be analyzed. Since displacement operators on disjoint sets of modes commute, the output displacement factorizes as
\begin{align}
    \hat D_{\rm out}(\mu')
    =
    \hat D_A(P_A\mu')\otimes \hat D_B(P_B\mu'),
\end{align}
where $P_A$ and $P_B$ denote the projections of the displacement vector onto the modes in $A$ and $B$, respectively. Therefore, $\hat D_{\rm out}(\mu')$ is a local unitary with respect to the bipartition $A:B$. It does not change the Schmidt coefficients, and hence does not change the R\'enyi entanglement entropy. Thus, it suffices to bound the R\'enyi entanglement entropy of the centered output state $|\Phi\rangle$.

\subsection{Output-state expansion and coefficient matrices}

We first expand the centered output state in the transformed Fock basis.
Writing $ |\psi_{\rm c}\rangle = \sum_{m=0}^{\infty} c_m|m\rangle$, the passive interferometer gives
\begin{align}
    |\Phi\rangle
    =
    \hat U
    \left(
    |\psi_{\rm c}\rangle^{\otimes N}
    \otimes
    |0\rangle^{\otimes(M-N)}
    \right)=
    \sum_{\vec n\in\mathbb{N}_0^N}
    c_{\vec n}
    |\vec n;U\rangle , \quad\text{where}\quad c_{\vec n}=\prod_{j=1}^{N}c_{n_j}.
\end{align}
Here, the $\{|\vec{n};U\rangle\}_{\vec{n} \in \mathbb{N}_0^N}$ consists of the input Fock states transformed by passive linear optics:
\begin{align}
    |\vec{n};U\rangle\equiv \hat{U}\left(|n_1,\dots,n_N\rangle\otimes |0\rangle^{\otimes (M-N)}\right)
    =\prod\limits_{j=1}^N \frac{1}{\sqrt{n_j!}} \left(\sum\limits_{k=1}^M U_{jk}^*\hat{a}_k^\dagger\right)^{n_j}|0\rangle^{\otimes M}.
\end{align}
Because $\hat U$ is unitary and the input Fock states are
orthonormal, the family
$\{|\vec n;U\rangle\}_{\vec n\in\mathbb N_0^N}$
is orthonormal.

To obtain the R\'enyi entropy, we have to obtain the Schmidt coefficients. The Fock component $|\vec{n};U \rangle$ can be decomposed across the bipartition by projections of the delocalized transformed creation operator onto the two subsystems:
\begin{align}
    \sum\limits_{k=1}^M U_{jk}^*\hat{a}_k^\dagger
    =
    \hat B_{A,j}^\dagger
    +
    \hat B_{B,j}^\dagger,\qquad\text{where}\qquad
    \hat B_{A,j}^\dagger
    =
    \sum_{k\in A}U_{jk}^*\hat a_k^\dagger
    \quad\text{and}
    \quad
    \hat B_{B,j}^\dagger
    =
    \sum_{k\in B}U_{jk}^*\hat a_k^\dagger.
\end{align}
The operators $\hat B_{A,j}^\dagger$ and $\hat B_{B,j}^\dagger$ are supported on different subsystems, and therefore commute. 

We next represent each transformed Fock component by its coefficient matrix across the bipartition $A:B$. 

Choose orthonormal bases
$\{|r\rangle_A\}$ and $\{|s\rangle_B\}$ and write
\begin{align}
    |\vec n;U\rangle
    =
    \sum_{r,s}
    (C_{\vec n})_{r,s}
    |r\rangle_A|s\rangle_B .
\end{align}
The coefficient matrix of the centered output state is then
\begin{align}
    M
    =
    \sum_{\vec n\in\mathbb N_0^N}
    c_{\vec n}C_{\vec n}.
\end{align}
Since the reduced centered output state is $\hat\rho_{{\rm c},A}\equiv\Tr_B(|\Phi\rangle\langle \Phi|)=MM^\dagger$, the Schmidt coefficients of the centered output state are the singular values of $M$, and hence $\Tr[\hat{\rho}_{\text{c},A}^\alpha]=\Vert M\Vert_{2\alpha}^{2\alpha}$.

Therefore, the R\'enyi entanglement entropy satisfies
\begin{align}
    S_\alpha(\hat{\rho}_A)=\frac{1}{1-\alpha}\log \Tr(\hat{\rho}_{\text{c},A}^\alpha) =\frac{1}{1-\alpha}\log \lVert M\rVert_{2\alpha}^{2\alpha}.
\end{align}
Therefore, we have to bound $\Vert M\Vert_{2\alpha}^{2\alpha}$.

\subsection{Bounding the R\'enyi entropy}

Before considering the infinite sum, we first establish the Schatten-norm bound for a finite set
$\Lambda\subset\mathbb N_0^N$.
Define $M_\Lambda \equiv \sum_{\vec n\in\Lambda} c_{\vec n}C_{\vec n}$. The key ingredient is an upper bound on the Schmidt rank of each transformed Fock component.

To establish this Schmidt-rank bound, consider an input mode $j$ with
occupation number $n_j$. Since $\hat B_{A,j}^\dagger$ and
$\hat B_{B,j}^\dagger$ commute, the corresponding factor admits the
exact binomial expansion
\begin{align}
    \left(
    \hat B_{A,j}^\dagger
    +
    \hat B_{B,j}^\dagger
    \right)^{n_j}
    =
    \sum_{r_j=0}^{n_j}
    \binom{n_j}{r_j}
    (\hat B_{A,j}^\dagger)^{r_j}
    (\hat B_{B,j}^\dagger)^{n_j-r_j}.
\end{align}
The binomial expansion then gives
\begin{align}
    |\vec n;U\rangle
    =
    \sum_{r_1=0}^{n_1}\cdots\sum_{r_N=0}^{n_N}
    \Gamma_{\vec r}
    \left[
        \prod_{j=1}^N(\hat{B}_{A,j}^\dagger)^{r_j}|0\rangle_A
    \right]
    \otimes
    \left[
        \prod_{j=1}^N(\hat B_{B,j}^\dagger)^{n_j-r_j}|0\rangle_B
    \right],
\quad\text{where}\quad
    \Gamma_{\vec r}
    =
    \prod_{j=1}^N
    \frac{\binom{n_j}{r_j}}{\sqrt{n_j!}}.
\end{align}
Thus, $|\vec n;U\rangle$ is expressed as a sum of at most $\prod_{j=1}^N(n_j+1)$ product vectors across $A:B$. These product vectors need not be normalized, orthogonal, or linearly independent.
Therefore,
\begin{align}
    \operatorname{SchmidtRank}(|\vec n;U\rangle)
    \le
    R_{\vec n},
    \qquad
    R_{\vec n}
    \equiv
    \prod_{j=1}^N(n_j+1).
\end{align}

For a finite set $\Lambda$, define the linear map $T_\Lambda(x)= \sum_{\vec n\in\Lambda} x_{\vec n}C_{\vec n}$.
The normalization and Hilbert--Schmidt orthogonality of the
states $|\vec n;U\rangle$ imply $\Vert C_{\vec{n}}\Vert_2=1$ and $\Vert x_{\vec{n}}C_{\vec{n}}\Vert_2=|x_{\vec{n}}|$. Moreover, since the Schmidt rank of $|\vec{n};U\rangle$ is at most $R_{\vec{n}}$, the rank of $C_{\vec{n}}$ is at most $R_{\vec{n}}$, and by the Cauchy-Schwarz inequality for its singular values, $\Vert C_{\vec{n}}\Vert_1\leq \sqrt{R_{\vec{n}}}$ and therefore $\Vert x_{\vec{n}}C_{\vec{n}}\Vert_1\leq |x_{\vec{n}}|\sqrt{R_{\vec{n}}}$. 

Therefore, the endpoint bounds for $T_\Lambda$ are
\begin{align}
    \Vert T_\Lambda(x)\Vert_1&=\left\Vert\sum\limits_{\vec{n}\in\Lambda}x_{\vec{n}}C_{\vec{n}}\right\Vert_1\leq \sum\limits_{\vec{n}\in\Lambda}\Vert x_{\vec{n}}C_{\vec{n}}\Vert_1\leq \sum\limits_{\vec{n}\in\Lambda}|x_{\vec{n}}|\sqrt{R_{\vec{n}}},\text{ and}\\
    \Vert T_\Lambda(x)\Vert_2^2&=\sum\limits_{\vec{n},\vec{m}\in \Lambda}x_{\vec{n}}^* x_{\vec{m}}\Tr(C_{\vec{n}}^\dagger C_{\vec{m}})=\sum\limits_{\vec{n}\in\Lambda}\left|x_{\vec{n}}\right|^2,
\end{align}
by Hilbert-Schmidt orthogonality. 

Thus, $T_\Lambda$ is a contraction from the weighted space $\ell_1(\sqrt{R_{\vec n}})$ to $S_1$ and an isometry from $\ell_2$ to $S_2$. For $1<p<2$, complex interpolation between these two endpoint
bounds, with $\frac{1}{p}= 1-\theta+\frac{\theta}{2}$, gives
\begin{align}
    \left\|
    \sum\limits_{\vec{n}\in \Lambda} x_{\vec{n}}C_{\vec{n}} \right\|_p^p
    \leq
    \sum_{\vec n\in\Lambda}
    \left(
    |x_{\vec n}|\sqrt{R_{\vec n}}
    \right)^{p(1-\theta)}
    |x_{\vec n}|^{p\theta}=
    \sum_{\vec n\in\Lambda}
    \left(
    |x_{\vec n}|\sqrt{R_{\vec n}}
    \right)^{2-p}
    |x_{\vec n}|^{2p-2}
    =
    \sum_{\vec n\in\Lambda}
    |x_{\vec n}|^p
    R_{\vec n}^{(2-p)/2}.
\end{align}

Taking $x_{\vec{n}}=c_{\vec{n}}$ and $p=2\alpha$, we obtain
\begin{align}
    \left\|
    \sum\limits_{\vec{n}\in \Lambda} c_{\vec{n}}C_{\vec{n}} \right\|_{2\alpha}^{2\alpha}=\|M_\Lambda\|_{2\alpha}^{2\alpha}
    \leq
    \sum_{\vec n\in\Lambda}
    |c_{\vec n}|^{2\alpha}
    R_{\vec n}^{1-\alpha}.
    \label{eq:finite schatten bound}
\end{align}

\subsection{Passage from finite truncations to the full expansion}

We now extend the finite-truncation bound to the full infinite expansion.
Define the single-mode weighted coefficient sum
\begin{align}
    A_\alpha(\eta)
    \equiv
    \sum_{m=0}^{\infty}
    |c_m|^{2\alpha}(m+1)^{1-\alpha}.
\end{align}
Because the input coefficients factorize over the $N$ modes, the corresponding multi-mode sum factorizes as
\begin{align}
    \sum\limits_{\vec{n}\in\mathbb{N}_0^N}|c_{\vec{n}}|^{2\alpha}R_{\vec{n}}^{1-\alpha}=\sum\limits_{n_1,\dots,n_N\geq 0}\prod\limits_{j=1}^N[|c_{n_j}|^{2\alpha}(n_j+1)^{1-\alpha}]=A_\alpha(\eta)^N.
\end{align}
We show that $A_\alpha(\eta)$ is finite and satisfies $A_\alpha(\eta)\leq 1+C_\alpha\eta^{2\alpha}<\infty$ for a constant $C_\alpha$.

We begin by expressing the Fock-basis coefficient $c_m$ in a form
suitable for bounding its absolute value:
\begin{align}\label{eq:coeff of fock basis}
    c_m
    =
    e^{-\mu^2/2}
    \frac{(-\mu)^m}{\sqrt{m!}}
    \left[
    \sqrt{1-\eta}
    +
    \sqrt{\eta}
    \left(
    \mu-\frac{m}{\mu}
    \right)
    \right]
    =
    e^{-\mu^2/2}
    \frac{(-\mu)^m}{\sqrt{m!}}
    \frac{1-\eta^2-m}{\sqrt{1-\eta}}.
\end{align}

Here, we can upper-bound the absolute value of each coefficient as follows:
\begin{align}
    |c_0|\leq 1,\qquad
    |c_1|
    =
    e^{-\mu^2/2}
    \mu
    \frac{\eta^2}{\sqrt{1-\eta}}
    =
    e^{-\mu^2/2}\eta^{5/2}
    \leq
    \eta^{5/2},
    \qquad
    \text{and}
    \qquad
    |c_m|
    \leq
    m\frac{\eta^{m/2}}{\sqrt{m!}},
    \qquad
    m\geq 2,
\end{align}
where we used $|1-\eta^2-m|\leq m$, and $\frac{\mu^m}{\sqrt{1-\eta}}=\eta^{m/2}(1-\eta)^{(m-1)/2}\leq \eta^{m/2}$.

We now bound $A_\alpha(\eta)$.
The $m=0$ contribution is bounded by $|c_0|^{2\alpha}\leq 1$.
The $m=1$ contribution satisfies
\begin{align}
    |c_1|^{2\alpha}2^{1-\alpha}
    \leq
    2^{1-\alpha}\eta^{5\alpha}
    \leq
    2^{1-\alpha}\eta^{2\alpha},
\end{align}
where we used $0<\eta< 1$ in the last inequality. For $m\geq 2$,
\begin{align}
    |c_m|^{2\alpha}(m+1)^{1-\alpha}
    \leq m^{2\alpha}(m+1)^{1-\alpha}
    \frac{\eta^{\alpha m}}{(m!)^\alpha}.
\end{align}
Since $m\geq 2$ and $\eta^{\alpha m}=\eta^{2\alpha}\eta^{\alpha(m-2)} \leq \eta^{2\alpha}$,
\begin{align}
    \sum_{m=2}^{\infty}
    |c_m|^{2\alpha}(m+1)^{1-\alpha}
    \leq C_\alpha'\eta^{2\alpha}, \quad\text{where}\quad
 C_\alpha'\equiv \sum_{m=2}^{\infty}  m^{2\alpha}\frac{(m+1)^{1-\alpha}}{(m!)^\alpha}  <\infty .
\end{align}

Combining the three contributions and redefining the constant
$C_\alpha$, we obtain $A_\alpha(\eta) \leq 1+2^{1-\alpha}\eta^{2\alpha} + C_\alpha'\eta^{2\alpha}= 1+C_\alpha\eta^{2\alpha}< \infty$. To pass to the full expansion, we choose an increasing sequence of finite sets $\{\Lambda_k\}_{k=1}^{\infty}$ of occupation-number vectors. Each set contains the preceding one, and every vector in $\mathbb N_0^N$ is eventually included.

We first show that $\{M_{\Lambda_k}\}_{k=1}^{\infty}$ is Cauchy in
$S_{2\alpha}$. Completeness then yields an $S_{2\alpha}$ limit $D$,
which will subsequently be identified with the full coefficient
matrix $M$ using convergence in $S_2$. For $k'>k$, the finite-truncation
bound gives
\begin{align}
\left\Vert M_{\Lambda_{k'}}-M_{\Lambda_k}\right\Vert_{2\alpha}^{2\alpha}\leq \sum\limits_{\vec{n}\in \Lambda_{k'}\backslash\Lambda_k}|c_{\vec{n}}|^{2\alpha}R_{\vec{n}}^{1-\alpha}\leq \sum\limits_{\vec{n}\notin \Lambda_k}|c_{\vec{n}}|^{2\alpha}R_{\vec{n}}^{1-\alpha}
\end{align}
and $\sum\limits_{\vec{n}}|c_{\vec{n}}|^{2\alpha}R_{\vec{n}}^{1-\alpha}=A_\alpha(\eta)^N<\infty$, the tail is $\sum\limits_{\vec{n}\notin \Lambda_k}|c_{\vec{n}}|^{2\alpha}R_{\vec{n}}^{1-\alpha}\rightarrow 0$. Therefore, the sequence $\{M_{\Lambda_k}\}$ is Cauchy in $S_{2\alpha}$. Since $S_{2\alpha}$ is complete, there exists $D\in S_{2\alpha}$ such that $M_{\Lambda_k}\rightarrow D$ in $S_{2\alpha}$.

Because $2\alpha<2$, monotonicity of Schatten norms gives
\begin{align}
\left\Vert M_{\Lambda_k}-D\right\Vert_2\leq \left\Vert M_{\Lambda_k}-D\right\Vert_{2\alpha}.
\end{align}
Therefore, in $S_2$, $M_{\Lambda_k}\rightarrow D$. 

Meanwhile, by Hilbert-Schmidt orthogonality of $C_{\vec{n}}$ and $\sum_{\vec{n}}|c_{\vec{n}}|^2=1$, 
\begin{align}
    \left\Vert M-M_{\Lambda_k}\right\Vert_2^2=\left\Vert\sum\limits_{\vec{n}\notin \Lambda_k}c_{\vec{n}}C_{\vec{n}}\right\Vert_2^2=\sum\limits_{\vec{n}\notin \Lambda_k}|c_{\vec{n}}|^2\rightarrow 0.
\end{align}

By uniqueness of the $S_2$ limit, $D=M
$. Therefore, the limit of $M_{\Lambda_k}$ in $S_{2\alpha}$ is also $M$. 

Taking $k\to\infty$ in Eq.~\eqref{eq:finite schatten bound} yields, 
\begin{align}
    \Vert M\Vert_{2\alpha}^{2\alpha}=\lim\limits_{k\rightarrow \infty }\left\Vert M_{\Lambda_k}\right\Vert_{2\alpha}^{2\alpha}\leq \lim\limits \sum\limits_{\vec{n}\in\Lambda_k}|c_{\vec{n}}|^{2\alpha} R_{\vec{n}}^{1-\alpha}=\sum\limits_{\vec{n}\in \mathbb{N}_0^N}|c_{\vec{n}}|^{2\alpha} R_{\vec{n}}^{1-\alpha}=A_{\alpha}(\eta)^N.\label{eq: upper bound of alpha norm}
\end{align}

Therefore, the R\'enyi entanglement entropy is upper bounded as 
\begin{align}
    S_\alpha(\hat{\rho}_A)=\frac{1}{1-\alpha}\log \Vert M\Vert_{2\alpha}^{2\alpha}\leq \frac{N}{1-\alpha}\log (1+C_\alpha\eta^{2\alpha})\leq \frac{C_\alpha}{1-\alpha}N\eta^{2\alpha}\equiv K_\alpha N\eta^{2\alpha}.
\end{align}

\section{Bound on R\'enyi entanglement entropy in Fock-state boson sampling}\label{appendix: ere in fock}

We now prove the coefficient bound required for the fixed-$\ell$ Fock-state extension in Sec.~\ref{sec:fock}. Because each sampled phase can be absorbed into the passive interferometer, it suffices to consider the branch $ |\psi(0)\rangle^{\otimes N}\otimes|0\rangle^{\otimes(M-N)}$. The coefficient-matrix, Schmidt-rank, interpolation, and infinite-truncation arguments are identical to those in Appendix~\ref{appendix: ere in spbs}; we therefore present only the input-dependent centered-state coefficient bound.

\subsection{Removing the coherent part by a local displacement}

Define the centered single-mode state which eliminates displacement from $|\psi(0)\rangle$ as $|\psi_{\rm c}\rangle$. 
When 
\begin{align}
    |\psi(0)\rangle\equiv \sum\limits_{k=0}^\ell \sqrt{\binom{\ell}{k}\eta^k (1-\eta)^{\ell-k}}|k\rangle\equiv \sum\limits_{k=0}^\ell a_k|k\rangle,
\end{align}
the displacement of the state is $\mu=\langle \hat{a}\rangle=\sqrt{\ell\eta}+O(\eta^{3/2})$,
and the centered single-mode state is
\begin{align}
    |\psi_{\rm c}\rangle&=\hat{D}(-\mu)|\psi(0)\rangle\equiv \sum\limits_{m=0}^\infty c_m |m\rangle.
\end{align}

\subsection{Single-mode coefficient bounds}
To derive coefficient-wise bounds, we first obtain an exact expression
for $c_m\equiv \langle m|\psi_{\rm c}\rangle$. For real $\mu$, the displacement operator can be written in the
normally ordered form $\hat D(-\mu) =e^{-\mu^2/2} e^{-\mu\hat a^\dagger} e^{\mu\hat a}$.
The actions of the two exponential operators on a Fock state are
\begin{align}
e^{\mu \hat{a}}|k\rangle&=\sum\limits_{r=0}^\infty \frac{\mu^r}{r!}\hat{a}^r|k\rangle=\sum\limits_{r=0}^k \frac{\mu^r}{r!}\sqrt{\frac{k!}{(k-r)!}}|k-r\rangle,\\
e^{-\mu\hat{a}^\dagger}|k-r\rangle&=\sum\limits_{s=0}^\infty \frac{(-\mu)^s}{s!}(\hat{a}^\dagger)^s |k-r\rangle=\sum\limits_{s=0}^\infty \frac{(-\mu)^s}{s!}\sqrt{\frac{(k-r+s)!}{(k-r)!}}|k-r+s\rangle,
\end{align}
Combining these two expressions gives
\begin{align}
    e^{-\mu\hat{a}^\dagger}e^{\mu\hat{a}}|k\rangle&=\sum\limits_{r=0}^k \sum\limits_{s=0}^\infty \frac{\mu^r(-\mu)^s}{r!s!}\sqrt{\frac{k!}{(k-r)!}}\sqrt{\frac{(k-r+s)!}{(k-r)!}}|k-r+s\rangle,\\
    \langle m|e^{-\mu\hat{a}^\dagger} e^{\mu\hat{a}}|k\rangle&=\sum\limits_{\substack{0\leq r \leq k\\ m-k+r\geq 0}}\frac{\mu^{2r+m-k}(-1)^{m-k+r}}{r!(m-k+r)!}\frac{\sqrt{k!m!}}{(k-r)!}
\end{align}
Using $|\psi(0)\rangle=\sum_{k=0}^{\ell}a_k|k\rangle$, we thus obtain
\begin{align}
    c_m&=e^{-\mu^2/2}\sum\limits_{k=0}^{\ell}a_k \sum\limits_{\substack{0\leq r \leq k\\ m-k+r\geq 0}} \frac{\mu^{2r+m-k}\sqrt{k!m!}(-1)^{m-k+r}}{r!(m-k+r)!(k-r)!}.\label{eq:form of c_m}
\end{align}

We next determine the dependence of each summand in
Eq.~\eqref{eq:form of c_m} on the transmission rate $\eta$.
For fixed $\ell$, the input coefficients satisfy
\begin{align}
    a_k=\sqrt{\binom{\ell}{k}}\eta^{k/2}(1-\eta)^{(\ell-k)/2}=O(\eta^{k/2}).
\end{align}
Therefore, the term in each summand satisfies
\begin{align}
    a_k\mu^{m-k+2r}=O(\eta^{k/2})\eta^{\frac{m}{2}-\frac{k}{2}+r}=O(\eta^{\frac{m}{2}+r}).
\end{align}
This termwise counting identifies the possible leading powers of
$\eta$. Cancellations between terms may increase the actual leading
order, as occurs for $c_1$ below.
\begin{align}
    c_1=e^{-\mu^2/2}[a_1-\mu a_0+O(\eta^{3/2})]=e^{-\mu^2/2}\left(\sqrt{\ell\eta}-\sqrt{\ell\eta}+O(\eta^{3/2})\right)=O(\eta^{3/2}).
\end{align}
Similarly, expanding the coefficient $c_2$ gives
\begin{align}
    c_2=e^{-\mu^2/2}\left[a_0 \frac{\mu^2\sqrt{2}}{2!}+a_1\left(-\mu\sqrt{2}+\frac{\mu^3\sqrt{2}}{2}\right)+a_2\left(1-2\mu^2+O(\mu^4)\right)\right]=O(\eta).
\end{align}

To control the remaining coefficients, for $m\geq 2$ we establish a bound of the form
$|c_m|\leq \eta^{m/2}K_{m,\ell}$, where the constants $K_{m,\ell}$ satisfy
$\sum_{m=2}^\infty K_{m,\ell}^{2\alpha}(m+1)^{1-\alpha}<\infty$.
To this end, using $|a_k|\leq\sqrt{\binom{\ell}{k}}\eta^{k/2}$,
the displacement amplitude $\mu$ can be bounded in terms of a constant
$H_\ell$ as follows:
\begin{align}
    \mu\equiv\langle \psi(0)|\hat{a}|\psi(0)\rangle=\sum\limits_{s=0}^{\ell-1}\sqrt{s+1}a_s a_{s+1}\leq \sum\limits_{s=0}^{\ell-1}\sqrt{(s+1)\binom{\ell}{s}\binom{\ell}{s+1}}\eta^{s+\frac{1}{2}}\leq H_\ell \sqrt{\eta},
\end{align}
    where $H_\ell\equiv \sum\limits_{s=0}^{\ell-1}\sqrt{(s+1)\binom{\ell}{s}\binom{\ell}{s+1}}\geq 1$.

Using Eq.~\eqref{eq:form of c_m}, we next derive an upper bound on
$|c_m|$. After the change of variables $q\equiv k-r$, the triangle inequality gives
\begin{align}
    |c_m|\leq \sum\limits_{k=0}^\ell |a_k|\sum\limits_{q=0}^{\min (k,m)}\frac{|\mu|^{m+k-2q}\sqrt{k!m!}}{(k-q)!(m-q)!q!}.
\end{align}
We know that $|a_k|\leq \sqrt{\binom{\ell}{k}}\eta^{k/2}$, and $|\mu|\leq H_\ell \sqrt{\eta}$, hence
\begin{align}
    |a_k||\mu|^{m+k-2q}\leq \sqrt{\binom{\ell}{k}}H_\ell^{m+k-2q}\sqrt{\eta}^{k+m+k-2q}\leq \sqrt{\binom{\ell}{k}}H_\ell^{m+\ell}\eta^{m/2}.
\end{align}
We substitute this to bound $|c_m|$:
\begin{align}
    |c_m|\leq \eta^{m/2} H_\ell^{m+\ell}\sum\limits_{k=0}^\ell \sqrt{\binom{\ell}{k}}\sqrt{k!}\sum\limits_{q=0}^{\min(k,m)}\frac{\sqrt{m!}}{(k-q)!(m-q)!q!}.
\end{align}
Since $q\leq k \leq \ell$, $\frac{\sqrt{m!}}{(m-q)!}=\frac{m!/(m-q)!}{\sqrt{m!}}\leq \frac{(m+1)^q}{\sqrt{m!}}\leq \frac{(m+1)^\ell}{\sqrt{m!}}$ ,
\begin{align}
    |c_m|\leq \eta^{m/2} H_\ell^{m+\ell}\frac{(m+1)^\ell}{\sqrt{m!}}\sum\limits_{k=0}^\ell \sqrt{\binom{\ell}{k}}\sqrt{k!}\sum\limits_{q=0}^{\min (k,\ell)}\frac{1}{(k-q)!q!}\leq    \eta^{m/2} H_\ell^{m+\ell}\frac{(m+1)^\ell}{\sqrt{m!}}\sum\limits_{k=0}^\ell \sqrt{\binom{\ell}{k}}\sqrt{k!}\sum\limits_{q=0}^{k}\frac{1}{(k-q)!q!}.
\end{align}
The second inequality holds because all summands in the $q$-sum are nonnegative. Using the identity $\sum\limits_{q=0}^k \frac{1}{(k-q)!q!}=\frac{2^k}{k!}$, 
\begin{align}
    |c_m|\leq \eta^{m/2} H_\ell^{m+\ell}\frac{(m+1)^\ell}{\sqrt{m!}}\sum\limits_{k=0}^\ell \sqrt{\binom{\ell}{k}}\sqrt{k!}\frac{2^k}{k!}=K_{m,\ell}\eta^{m/2},\quad\text{where}\quad K_{m,\ell}\equiv \frac{H_\ell^{m+\ell}(m+1)^\ell}{\sqrt{(m!)}}\sum\limits_{k=0}^\ell \sqrt{\binom{\ell}{k}}\frac{2^k}{\sqrt{k!}}.
\end{align}

For $m\geq 2$, the sum is
\begin{align}
    \sum\limits_{m=2}^\infty |c_m|^{2\alpha}(m+1)^{1-\alpha}\leq \sum\limits_{m=2}^\infty \eta^{\alpha m}K_{m,\ell}^{2\alpha}(m+1)^{1-\alpha}\leq C_{\alpha,\ell}' \eta^{2\alpha}.
\end{align}

Here, we define the constant 
\begin{align}
    C_{\alpha,\ell}'\equiv \sum\limits_{m=2}^\infty K_{m,\ell}^{2\alpha}(m+1)^{1-\alpha}=\sum\limits_{m=2}^\infty \frac{H_\ell^{2\alpha (m+\ell)}(m+1)^{2\alpha \ell+1-\alpha}}{(m!)^\alpha}\left[\sum\limits_{k=0}^\ell \sqrt{\binom{\ell}{k}}\frac{2^k}{\sqrt{k!}}\right]^{2\alpha}<\infty. 
\end{align}

Combining the bounds for $m=0$, $m=1$, and $m\geq2$ gives
\begin{align}
    A_{\alpha,\ell}(\eta)
    \leq
    1+C_{\alpha,\ell}\eta^{2\alpha}.
\end{align}
Here, normalization gives $|c_0|\leq1$, while the
$m=1$ contribution is of higher order in $\eta$ and therefore
does not affect the leading $\eta^{2\alpha}$ scaling.

\section{Bound on R\'enyi entanglement entropy in cat state boson sampling}\label{appendix: ere in cat}

As in Appendices~\ref{appendix: ere in spbs} and~\ref{appendix: ere in fock}, our goal is to obtain the upper bound on $\Tr(\hat{\rho}_A^\alpha)$ using $A_{\alpha,\gamma}(\eta)=\sum\limits_{m=0}^\infty |c_m|^{2\alpha}(m+1)^{1-\alpha}\leq 1+C_{\alpha,\gamma}\eta^{2\alpha}$.

\subsection{Removing the coherent part by a local displacement}
We now prove the single-mode coefficient bound required for the fixed-amplitude cat-state extension.
Consider the lossy odd-cat input $\hat\sigma^{\otimes N}
    \otimes
    |0\rangle\langle0|^{\otimes(M-N)}$,
with the decomposition in
Eq.~\eqref{eq:lossy cat state decomposition}.
Because
$|\psi_-\rangle=e^{i\pi\hat n}|\psi_+\rangle$,
the two branches differ only by an input phase shift.
It therefore suffices to analyze the all-$+$ branch $|\psi_+\rangle^{\otimes N}\otimes |0\rangle^{\otimes(M-N)}$.
We also take fixed $\gamma>0$ to be real, since its phase can be
absorbed into the passive interferometer.

The displacement of the state is
    $\mu=\langle \psi_+|\hat{a}|\psi_+\rangle=\gamma\sqrt{\eta} (a_+^2-a_-^2)=r\gamma\sqrt{\eta}$, 
where $r\equiv \frac{\sqrt{1-e^{-4\gamma^2(1-\eta)}}}{1-e^{-2\gamma^2}}$. The range of $r$ is 
\begin{align}
0\leq r \leq \frac{\sqrt{1-e^{-4\gamma^2}}}{1-e^{-2\gamma^2}}\leq \sqrt{\frac{1+e^{-2\gamma^2}}{1-e^{-2\gamma^2}}}\equiv R_\gamma.    
\end{align}
We can upper bound $|\mu|=|r\gamma\sqrt{\eta}|\leq R_\gamma\gamma\sqrt{\eta}$.

The centered single-mode state after removing the displacement, which does not change the Schmidt spectrum, is 
\begin{align}
|\psi_{\rm c}\rangle=\hat{D}(-\mu)|\psi_+\rangle=a_+|\gamma\sqrt{\eta}-\mu\rangle-a_-|-\gamma\sqrt{\eta}-\mu\rangle\equiv \sum\limits_{m=0}^\infty c_m|m\rangle.
\end{align}

\subsection{Single-mode coefficient bounds}

The coefficient of $|\psi_{\rm c}\rangle$ in the Fock basis $|m\rangle$ is 
\begin{align}
    c_m=\langle m|\psi_{\rm c}\rangle= a_+e^{-\gamma^2\eta(1-r)^2/2}\frac{(\gamma\sqrt{\eta})^m(1-r)^m}{\sqrt{m!}}-a_-e^{-\gamma^2\eta(1+r)^2/2}\frac{(-\gamma\sqrt{\eta})^m (1+r)^m}{\sqrt{m!}}.
\end{align}

We can first bound the coefficient in the Fock basis. $|c_0|\leq 1$ because of the normalization condition, and the leading term of $c_1$ is

\begin{align}
    c_1&=a_+e^{-\gamma^2\eta(1-r)^2/2}\gamma\sqrt{\eta}(1-r)-a_-e^{-\gamma^2\eta(1+r)^2/2}(-\gamma\sqrt{\eta})(1+r)\\
    &=\gamma\sqrt{\eta}
    \left[
    a_+(1-r)+a_-(1+r)
    \right]+
    \gamma\sqrt{\eta}
    \left[
    a_+(1-r)
    \left(
    e^{-\frac{1}{2}\gamma^2\eta(1-r)^2}-1
    \right)
    +
    a_-(1+r)
    \left(
    e^{-\frac{1}{2}\gamma^2\eta(1+r)^2}-1
    \right)
    \right].
\end{align}

The order of the first term is
\begin{align}
    \gamma\sqrt{\eta}[(a_+ +a_-)-r(a_+-a_-)]=\gamma\sqrt{\eta}\frac{\sqrt{1+e^{-2\gamma^2(1-\eta)}}(e^{-2\gamma^2(1-\eta)}-e^{-2\gamma^2})}{(1-e^{-2\gamma^2})^{3/2}}\leq \gamma\sqrt{\eta}\frac{\sqrt{2}(1-e^{-2\gamma^2})\eta}{(1-e^{-2\gamma^2})^{3/2}}=\gamma\sqrt{\frac{2}{1-e^{-2\gamma^2}}}\eta^{3/2}.
\end{align}
We used $e^{-2\gamma^2(1-\eta)}\leq (1-\eta)e^{-2\gamma^2}+\eta$ due to convexity of $e^{-2\gamma^2(1-\eta)}$.

Using $|e^{-t}-1|\leq t$ and $|1\pm r|\leq 1+R_\gamma$, the second term is less than or equal to $\frac{\gamma^3}{2}\sqrt{\frac{2}{1-e^{-2\gamma^2}}}\eta^{3/2}(1+R_\gamma)^3$.

Therefore, 
\begin{align}
    |c_1|\leq \gamma\sqrt{\frac{2}{1-e^{-2\gamma^2}}}\left[1+\frac{\gamma^2}{2}(1+R_\gamma)^3\right]\eta^{3/2}.
\end{align}

The absolute value of $c_m$ when $m\geq 2$ is upper-bounded by
\begin{align}
    |c_m|\leq |a_++a_-|\left(\frac{(\gamma\sqrt{\eta})^m(1+r)^m}{\sqrt{m!}}\right)\leq\frac{S_\gamma^m}{\sqrt{m!}}\eta^{m/2},\quad\text{where}\quad S_\gamma&\equiv \sqrt{\frac{2}{1-e^{-2\gamma^2}}}(1+R_\gamma)\gamma. 
    \label{eq: upper bound of c_m}
\end{align}

We next bound $A_{\alpha,\gamma}(\eta)$, whose $N$th power bounds $\Tr(\hat{\rho}_A^\alpha)$, as in Eq.~\eqref{eq: upper bound of alpha norm}. When $m\geq 2$, for fixed $\gamma>0$,
\begin{align}
\sum\limits_{m=2}^\infty \left\vert c_m\right\vert^{2\alpha}(1+m)^{1-\alpha}\leq \sum\limits_{m=2}^\infty \frac{S_\gamma^{2m\alpha}\eta^{m\alpha}(1+m)^{1-\alpha}}{(m!)^\alpha}\leq C_{\alpha,\gamma}'\eta^{2\alpha},\quad\text{where}\quad 
C_{\alpha,\gamma}'\equiv \sum\limits_{m=2}^\infty \frac{S_\gamma^{2\alpha m}(m+1)^{1-\alpha}}{(m!)^\alpha}<\infty.
\end{align}

Combining the three contributions gives
\begin{align}
    A_{\alpha,\gamma}(\eta)
    \leq 1+\left\vert c_1\right\vert^{2\alpha}2^{1-\alpha}+C_{\alpha,\gamma}'\eta^{2\alpha}\leq 1+C_{\alpha,\gamma}\eta^{2\alpha}.
\end{align}
The $m=1$ contribution is of higher order in $\eta$; the leading correction arises from the $m=2$ sector.
Applying the coefficient-matrix and Schatten-norm argument of Appendix~\ref{appendix: ere in spbs} then gives
\begin{align}
    S_\alpha(\hat\rho_A)
\le
\frac{N}{1-\alpha}
\log\left(1+C_{\alpha,\gamma}\eta^{2\alpha}\right)
\le
K_{\alpha,\gamma}N\eta^{2\alpha}.
\end{align}

We finally verify that the infinite Fock expansion does not introduce an additional superpolynomial computational overhead. For fixed $\gamma>0$, Eq.~\eqref{eq: upper bound of c_m} gives
\begin{align}
    \sum_{m=d+1}^{\infty}|c_m|^2
    \leq
    \sum_{m=d+1}^{\infty}
    \frac{S_\gamma^{2m}\eta^m}{m!}
    \leq
    \sum_{m=d+1}^{\infty}
    \frac{S_\gamma^{2m}}{m!},
\end{align}
where $S_\gamma<\infty$ is independent of $N$ and $\eta$. For $d$ sufficiently large compared with the $\gamma$-dependent constant $S_\gamma^2$, the ratio between successive terms in the last series is at most $1/2$. It follows that
\begin{align}
    \sum_{m=d+1}^{\infty}|c_m|^2
    \leq
    2\frac{S_\gamma^{2(d+1)}}{(d+1)!}
    \leq
    2\left(
    \frac{eS_\gamma^2}{d+1}
    \right)^{d+1},
\end{align}
where we used
$(d+1)!\geq[(d+1)/e]^{d+1}$ and we take $d$ satisfying $d+1\geq 2eS_\gamma^2$.

For the $N$-mode product input, the total Fock weight discarded by restricting every centered input mode to occupation numbers $0,\ldots,d$ is
\begin{align}
    1-\left(\sum_{m=0}^{d}|c_m|^2
    \right)^N=1-\left( 1-\sum_{m=d+1}^{\infty}|c_m|^2
    \right)^N\leq N\sum_{m=d+1}^{\infty}|c_m|^2\leq N2^{-d}.
\end{align}
Consequently, choosing $d= O\left(\log\frac{N}{\varepsilon} \right)$ makes the total discarded Fock weight at most $\varepsilon$.
After this truncation, the total photon number is at most $Nd$.
Since passive linear optics preserves the total photon number, an
occupation-number cutoff of $Nd$ on each physical output mode is
sufficient. The corresponding local Hilbert-space dimension is therefore
at most $Nd+1=O\left( N\log\frac{N}{\varepsilon} \right)$, which is polynomially bounded.

\section{Analysis of the required bond dimension for classical simulation of boson sampling}\label{appendix: numerical bond dimension analysis}

\subsection{Approximate canonical-mode estimate}~\label{appendix: subsection bond dimension analysis algorithm}

This Appendix describes the approximate canonical-mode procedure used to generate the boson-sampling resource estimates in Fig.~\ref{fig:numerical_analysis}.
This approximation is independent of the rigorous R\'enyi-entropy bound proved in
Appendix~\ref{appendix: ere in spbs} and is used only for finite-size numerical resource estimation in Sec.~\ref{section: numerical computational cost}.

To analyze the entanglement of the output state, we consider the balanced spatial bipartition $A=\{1,\ldots,l\}$ and $B=\{l+1,\ldots,M\}$ of the physical output modes, with
$l=\lfloor M/2\rfloor$. We write the unitarily transformed creation operator as
\begin{align}
    \sum\limits_{k=1}^M U_{jk}^*\hat{a}_k^\dagger
    \equiv\cos\theta_j\hat{B}_{A,j}^\dagger+\sin\theta_j\hat{B}_{B,j}^\dagger,
\end{align}
where we define 
\begin{align}
    \cos\theta_j\hat{B}_{A,j}^\dagger=\sum\limits_{k=1}^l U_{jk}^*\hat{a}_k^\dagger,\qquad \sin\theta_j\hat{B}_{B,j}^\dagger =\sum\limits_{k=l+1}^M U_{jk}^*\hat{a}_k^\dagger. 
\end{align}
In this setting, $\cos^2\theta_j$ is the probability that a photon initially occupying input mode $j$ is detected in the physical output subsystem $A$, while $\sin^2\theta_j$ is the corresponding probability for subsystem $B$.
By definition, the normalization is given as
\begin{align}
\cos^2\theta_j=\sum\limits_{k=1}^l|U_{jk}|^2,\qquad \sin^2\theta_j=\sum\limits_{k=l+1}^M |U_{jk}|^2.
\end{align}

Unlike the analytical treatment in Sec.~\ref{subsection: ere of the output state}, we assume that the operators $\hat{B}_{A,j}^\dagger$ and $\hat{B}_{B,j}^\dagger$ follow the canonical commutation relation approximately, 
\begin{align}
&\left[\hat{B}_{A,j},\hat{B}_{A,k}^\dagger\right]\approx\delta_{jk},~~~ \left[\hat{B}_{B,j},\hat{B}_{B,k}^\dagger\right]\approx\delta_{jk}, ~~~\left[\hat{B}_{A,j},\hat{B}_{B,k}^\dagger\right]=0,~~~\left[\hat{B}_{A,j}^\dagger,\hat{B}_{B,k}\right]=0. \label{eq:canonical}
\end{align}

Using these definitions, we now evaluate the R\'enyi entanglement entropy between bipartite subsystems of equal size; therefore, we set $ l = \lfloor M/2 \rfloor$.
To do that, we now compute the eigenvalues of the reduced density matrix on the subsystem $A$ (or equivalently the subsystem $B$).

Substituting these $\hat{B}$ operators transforms the output state $|\psi_{\text{out}}\rangle$ as
\begin{align}
    &\bigotimes\limits_{j=1}^N \left[ \left(\sqrt{1-\eta}+\sqrt{\eta}\left(\cos\theta_j\hat{B}_{A,j}^\dagger+\sin\theta_j\hat{B}_{B,j}^\dagger\right)\right)|0\rangle\right]=\bigotimes\limits_{j=1}^N\left[\sqrt{1-\eta}|00\rangle_j+\sqrt{\eta}\cos\theta_j|10\rangle_j+\sqrt{\eta}\sin\theta_j|01\rangle_j\right]. 
\end{align}
The remaining $M-N$ input modes are in the vacuum and therefore do
not generate additional factors in this auxiliary transformed-mode representation. They therefore do not affect the entanglement and are omitted. For brevity, we also omit the labels of subsystem $A$ and $B$  when it is clear from the context; for example, we write $\hat{B}_{A,j}^\dagger|0\rangle=|1\rangle_{A,j}|0\rangle_{B,j}$ simply as $|10\rangle_j$, where the first and second entries refer to subsystems $A$ and $B$, respectively.

To derive the R\'enyi entanglement entropy across the bipartition, we find the eigenvalues of the reduced density matrix of $|\psi_{\text{out}}\rangle\langle\psi_{\text{out}}|$. After tracing out the subsystem $B$, the reduced density matrix on the subsystem $A$ can be represented as the following matrix:
\begin{align}
    \label{eq:EV of BS}
    \bigotimes_{j=1}^N \begin{pmatrix}1-\eta \cos^2\theta_j&\sqrt{\eta(1-\eta)}\cos\theta_j\\ \sqrt{\eta(1-\eta)}\cos\theta_j&\eta\cos^2\theta_j\end{pmatrix},
\end{align}
where we have omitted the trivial tensor product $|0\rangle\langle 0|^{\otimes(M-N)}$ as above.
Therefore, the eigenvalues of the reduced density matrix are given as 
\begin{align}
    p(\vec{k})
    \equiv \prod_{j=1}^N c_{jk_j}=\prod_{j=1}^N c_{j0}^{1-k_j} c_{j1}^{k_j}.
\end{align}
where $\vec{k}\in\{0,1\}^N$ and
\begin{align}
c_{j0}=\dfrac{1}{4}\left(2+\sqrt{2\eta^2\cos(4\theta_j)-2\eta^2+4}\right),~~c_{j1}=\dfrac{1}{4}\left(2-\sqrt{2\eta^2\cos(4\theta_j)-2\eta^2+4}\right).
\end{align}

Hence,
\begin{align}
    \log p(\vec{k})&=\sum\limits_{j=1}^N \left[(1-k_j)\log c_{j0}+k_j\log c_{j1}\right]=\sum\limits_{j=1}^N \log c_{j0}+\sum\limits_{j=1}^N k_j\log(\dfrac{c_{j1}}{c_{j0}})=C+\sum\limits_{j=1}^N k_j\log r_j,
\label{eq:log_probability}
\end{align}
where we define $C\equiv\sum_{j=1}^N \log c_{j0}$ and $r_j\equiv c_{j1}/c_{j0}<1$. 
Note that as $r_j$ decreases, $p(\vec{k})$ becomes a smaller eigenvalue and thus less significant. 

We construct the list of $\log p(\vec{k})$ iteratively. 
Initialize $L=\{0\}$, and for each $j$ update
\begin{align}
  L \leftarrow L \,\cup\, \{\;\ell + \log r_j \;:\; \ell \in L\;\}.
\end{align}
Here, when appending elements to the list, all eigenvalue multiplicities are retained. Since $\log r_j < 0$, we can prune partial sums during the construction:
for any $t < N$, if a partial value $\sum_{j=1}^t k_j^* \log r_j$ is below the threshold, then every $\vec{k}$ whose first $t$ entries coincides with $\vec{k}^*$ will also be below the threshold; hence we discard that entire branch from $L$. 
This pruning removes eigenvalues below the threshold.

By choosing the threshold so that the cumulative weight of the retained eigenvalues is sufficiently close to the target accuracy, say $99\%$, and then counting the eigenvalues that exceed the threshold, we obtain an estimate of the required bond dimension. 

The code is available at https://doi.org/10.5281/zenodo.21149576.
\subsection{Validation against exact calculation}
~\label{appendix: subsection comparison between exact and est case}

\begin{figure}[h]
    \centering
    \includegraphics[width=0.5\linewidth]{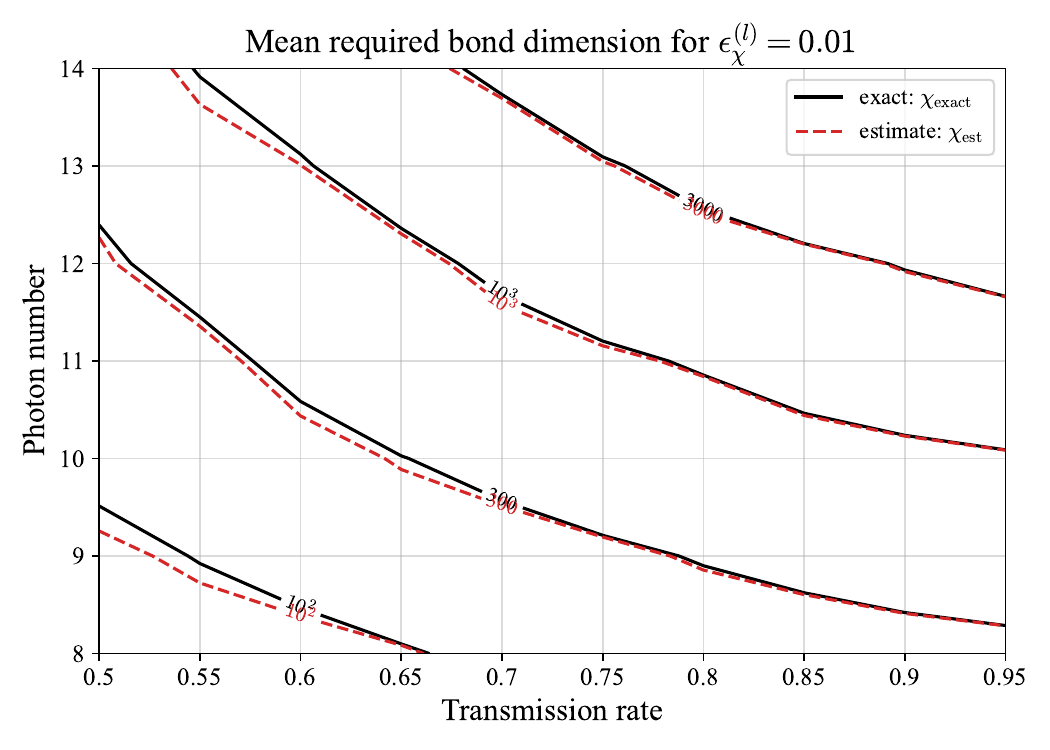}
    \caption{The mean required bond dimension of the exact calculation and the approximate canonical-mode model, with respect to photon number and the transmission rate of the circuit. The contour represents the required bond dimension of each case. The unitary of the circuit is assumed to be Haar-random, and the local discarded-weight threshold is set to be $0.01$. 
     }
    \label{fig: chi_exact and est_contour}
\end{figure}

We next compare the approximate canonical-mode estimate with an exact finite-size calculation without the approximate canonical-mode assumption.
Exact calculations become costly as $N$ increases, so we restrict the comparison to $N\leq 14$.
For Haar-random interferometers, Fig.~\ref{fig: chi_exact and est_contour} compares the
resulting bond-dimension contours.
Within the tested range, the two calculations show consistent behavior, with the approximate estimate being slightly larger for the instances considered.
We therefore use it as an empirically conservative proxy in the resource plots, without claiming that this ordering holds for all system sizes or interferometers.

\end{document}